\documentclass[12pt,titlepage]{article}

\newif\ifdraft
\ifx\draftmode\undefined
  \draftfalse
\else
  \drafttrue
\fi

\usepackage[pdftex]{graphicx}
\DeclareGraphicsExtensions{.jpg,.pdf,.eps,.png,.tiff}

\ifdraft
  
\else
  \usepackage[nolists,nomarkers]{endfloat}
  \AtBeginDelayedFloats{}
\fi

\usepackage{helvet}

\usepackage{amssymb}

\usepackage{url}

\usepackage{enumitem}

\usepackage{cite}
\let\citeleft=(
\let\citeright=)

\usepackage{graphicx}
\usepackage{caption}
\usepackage{subcaption}

\usepackage{mathtools}

\usepackage{multirow}
\usepackage{tabularx}
\usepackage{array, booktabs}
\usepackage{makecell}
\usepackage[T1]{fontenc}

\newtagform{brackets}{[}{]}
\usetagform{brackets}

\graphicspath{{figs/}}

\begin{document}
\bibliographystyle{mrm}

\begin{titlepage}

\begin{center}
\textbf{\Large A Metabolite Specific 3D Stack-of-Spiral bSSFP Sequence for Improved Lactate Imaging in Hyperpolarized [1-$^{13}$C]Pyruvate Studies on a 3T Clinical Scanner}
\end{center}
\bigskip

\begin{center}
Shuyu Tang$^{1,2}$, Robert Bok$^{2}$, Hecong Qin$^{1,2}$, Galen Reed$^{3}$, Mark VanCriekinge$^{2}$, Romelyn Delos Santos$^{2}$, William Overall$^{2}$, Juan Santos$^{2}$, Jeremy Gordon$^{2}$,   \\
Zhen Jane Wang $^{1,2}$, Daniel B. Vigneron$^{1,2}$, Peder E.Z. Larson$^{1,2}$
\end{center}

\vspace*{0.1in}
\noindent
\noindent
$^1$UC Berkeley-UCSF Graduate Program in Bioengineering, University of California, San Francisco and University of California, Berkeley

\noindent
$^2$Department of Radiology and Biomedical Imaging,
University of California - San Francisco, San Francisco, California.  

\noindent
$^3$HeartVista, Los Altos, California, USA

\noindent
\noindent
{\em Address correspondence to:} \\
	Shuyu Tang \\
	Byers Hall, Room 102 \\
	1700 4th St \\
	San Francisco, CA \ 94158 \\
E-MAIL: shuyu.tang@ucsf.edu 

\noindent

\noindent
Approximate Word Count: 200 (abstract)  4692  (body) \\

\noindent
Submitted Sep 1, 2019, to {\it Magnetic Resonance in Medicine} as a Full Paper.\\

\end{titlepage}

\section*{Abstract}
\setlength{\parindent}{0in}


Purpose: The balanced steady-state free precession sequence has been previously explored to improve the efficient use of non-recoverable hyperpolarized $^{13}$C magnetization, but suffers from poor spectral selectivity and long acquisition time. The purpose of this study was to develop a novel metabolite-specific 3D bSSFP ("MS-3DSSFP") sequence with stack-of-spiral readouts for improved lactate imaging in hyperpolarized [1-$^{13}$C]pyruvate studies on a clinical 3T scanner.

Methods: Simulations were performed to evaluate the spectral response of the MS-3DSSFP sequence. Thermal $^{13}$C phantom experiments were performed to validate the MS-3DSSFP sequence. In vivo hyperpolarized [1-$^{13}$C]pyruvate studies were performed to compare the MS-3DSSFP sequence with metabolite specific gradient echo ("MS-GRE") sequences for lactate imaging. 

Results: Simulations, phantom and in vivo studies demonstrate that the MS-3DSSFP sequence achieved spectrally selective excitation on lactate while minimally perturbing other metabolites. Compared with MS-GRE sequences, the MS-3DSSFP sequence showed approximately a 2.5-fold SNR improvement for lactate imaging in rat kidneys, prostate tumors in a mouse model and human kidneys. 

Conclusions: Improved lactate imaging using the MS-3DSSFP sequence in hyperpolarized [1-$^{13}$C]pyruvate studies was demonstrated in animals and humans. The MS-3DSSFP sequence could be applied for other clinical applications such as in the brain or adapted for imaging other metabolites such as pyruvate and bicarbonate. 

\vspace{0.4in}
\setlength{\parindent}{0in}
{\bf Key words: bSSFP, $^{13}$C, Hyperpolarized, MRI, metabolic imaging, Pyruvate, Lactate}
\newpage

%
%
%
%
%

%
%
\section*{Introduction}
%
Magnetic resonance imaging with hyperpolarized $^{13}$C-labeled compounds via dynamic nuclear polarization (DNP) has been used to non-invasively study metabolic processes in vivo \cite{larsen2003a, golman2006a}. This method provides a transient signal enhancement of more than 10,000 fold compared to imaging $^{13}$C compounds at thermal equilibrium. However, as soon as the pre-polarized $^{13}$C-labeled compound leaves the polarizer, its hyperpolarized state irreversibly decay to the thermal equilibrium with a decay constant characterized by T1, which is typically less than one minute. Therefore, imaging approaches with fast readouts and efficient use of hyperpolarized magnetization are favorable. 

Hyperpolarized [1-$^{13}$C]pyruvate has been widely used to monitor metabolic pathways in a number of applications \cite{day2007a,albers2008a, schroeder2008a, park2010a, witney2010a, darpolor2011a} and its feasibility for clinical applications has been demonstrated \cite{nelson2013a, cunningham2016a, miloushev2018a, grist2019a}. The MR signals of the hyperpolarized [1-$^{13}$C]pyruvate (173 ppm) and its downstream metabolites - [1-$^{13}$C]lactate (185 ppm), [1-$^{13}$C]pyruvate hydrate (181 ppm), [1-$^{13}$C]bicarbonate (163 ppm) and [1-$^{13}$C]alanine (178 ppm) - are typically acquired using gradient echo ("GRE") sequences (CSI\cite{golman2006a, mayer2010a}, multi-echo IDEAL\cite{reeder2007a, wiesinger2012a}, metabolite specific EPI\cite{cunningham2008a, gordon2017a} or spiral\cite{lau2011a} acquisition) where the transverse magnetization is spoiled at the end of each repetition time.  Compared to GRE acquisitions, the balanced steady state free precession ("bSSFP") \cite{svensson2003a, leupold2009a, morze2013a,  reed2014a, milshteyn2017a, milshteyn2018a, shang2017a} sequence can acquire the nonrenewable hyperpolarized magnetization more efficiently by repetitively refocusing transverse spins, which is especially valuable for imaging metabolites with long T2s \cite{yen2010a, milshteyn2017a} such as [1-$^{13}$C]pyruvate or [1-$^{13}$C]lactate.

Our work focuses on improving lactate imaging with hyperpolarized [1-$^{13}$C]pyruvate injections using a bSSFP framework. Three bSSFP strategies for lactate imaging have been proposed in the prior works. The first strategy \cite{leupold2009a} utilized a broadband pulse to excite all components (i.e. pyruvate, lactate, bicarbonate, alanine, pyruvate-hydrate) in [1-$^{13}$C]pyruvate studies and decomposed the spectral information using multi-echo readouts. By acquiring all compounds at one time, this strategy limits the acquisition optimization (e.g. flip angle, resolution) for individual metabolites and would require a longer acquisition time if not all the compounds in the spectrum are of interest.

The second strategy \cite{milshteyn2018a} reduced the number of excited compounds - only excited lactate, pyruvate hydrate and alanine - and applied a saturation pulse to suppress undesired signals from alanine and pyruvate hydrate at the beginning of each bSSFP acquisition. There are three main drawbacks in this strategy. Since the conversion between pyruvate hydrate and pyruvate maintains an equilibrium in the liquid state \cite{pocker1969a,larson2012a}, the pre-saturated pyruvate hydrate signal may recover and still contaminate lactate acquisitions at later bSSFP echoes. Directly saturating pyruvate hydrate would also accelerate the loss of pyruvate magnetization and reduce the signals of downstream metabolites. In addition, the saturation performance may be imperfect in the regions where transmit B1 profile is not homogeneous.

The third strategy \cite{shang2017a} excited one metabolite at a time (i.e. metabolite specific excitation) without a suppression pulse and was applied for imaging [1-$^{13}$C]urea, [1-$^{13}$C]pyruvate and [1-$^{13}$C]lactate with the bSSFP sequence on a 14.1T scanner. To meet the constraint of short TR in bSSFP sequence, this strategy designed a multiband RF pulse using a convex optimization approach \cite{shang2016a}. Compared with single-band RF pulses, multiband RF pulses could potentially shorten the RF duration by releasing the constraints on frequency ranges of no interest. Our work adapted this strategy to a clinical 3T scanner.

This article presents a novel metabolite specific 3D bSSFP sequence ("MS-3DSSFP") with stack-of-spiral readouts for improved dynamic lactate imaging in hyperpolarized [1-$^{13}$C]pyruvate studies on a clinical 3T scanner. A lactate specific excitation pulse was developed using a previously described approach \cite{shang2016a} and stack-of-spiral readouts were used to accelerate the acquisition. The excitation profile of the newly designed RF pulse at the bSSFP state was simulated to investigate the banding artifacts and to examine the spectral selectivity of the RF pulse. Thermally polarized $^{13}$C phantom experiments were performed to validate the MS-3DSSFP sequence. In vivo hyperpolarized [1-$^{13}$C]pyruvate experiments were performed on healthy rats, prostate cancer mouse model and patients with renal tumors to compare the MS-3DSSFP sequence with metabolite specific GRE ("MS-GRE") sequences, in the aspects of signal-to-noise ratio (SNR), image artifacts and impact on other metabolites. 


\section*{Methods}
\subsubsection*{Sequence design and simulation}
\begin{figure} 
\centering  
\includegraphics[width=1\textwidth]{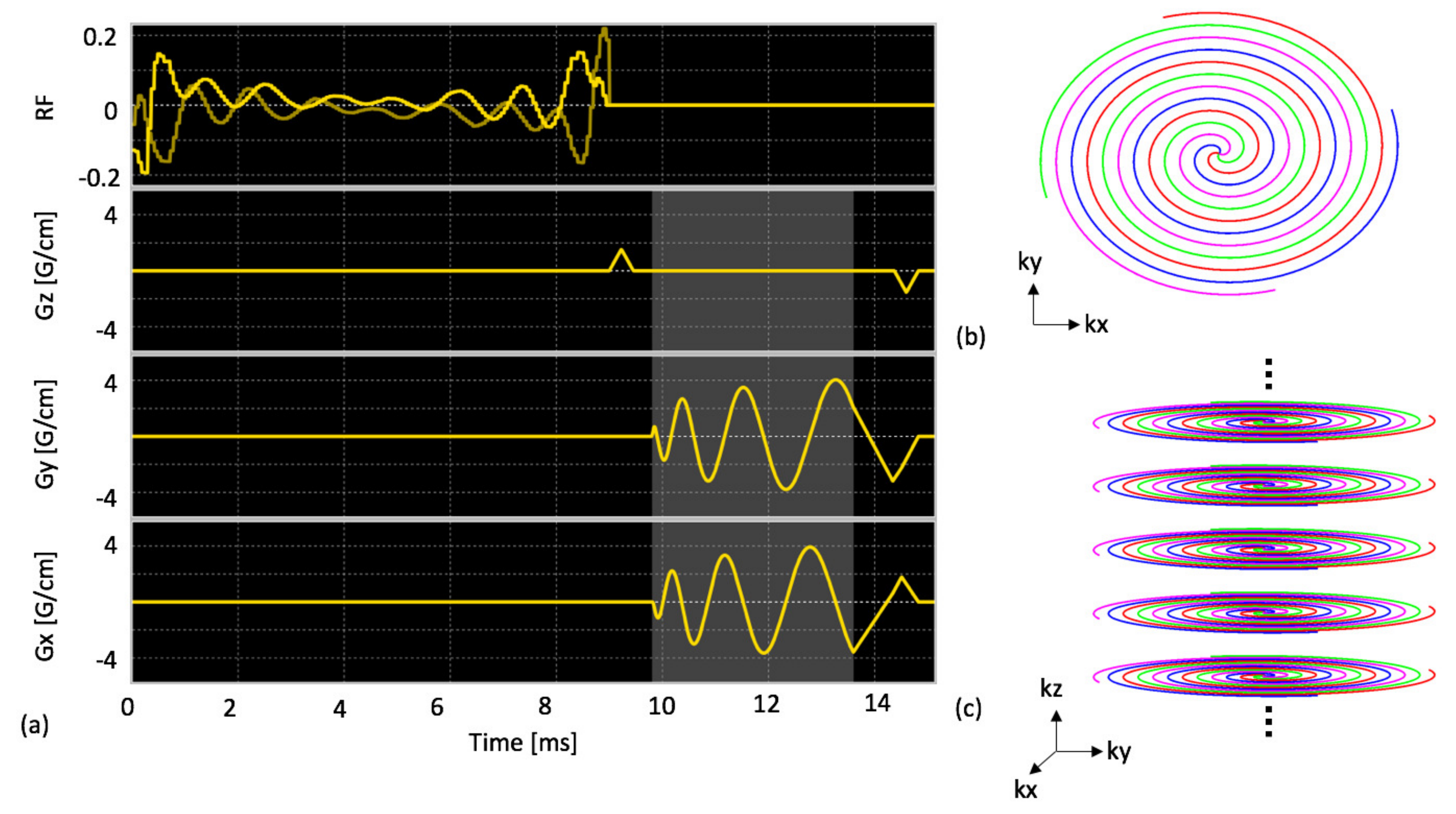}
 \caption
    {
Pulse sequence of the proposed MS-3DSSFP acquisition (a). It consists of a lactate specific excitation pulse and a 3D center-out stack of spiral readout (c). Each stack (b) consists of four interleaves. The details of the excitation pulse are described in Figure \ref{fig:excitation_profile}. Lighted shaded regions refer to the data acquisition window.
    }
  \label{fig:seq_scheme}
\end{figure}

The MS-3DSSFP sequence (Figure \ref{fig:seq_scheme}) consists of a multiband RF pulse and a center-out 3D uniform-density stack-of-spiral readout. The RF pulse was designed using a prior approach \cite{shang2016a} to minimize the pulse duration. This pulse had a duration of 9ms, a maximum B1 of 0.2195G, a 40Hz passband on lactate (0Hz), a 40Hz stopband with 5\% ripples on pyruvate hydrate (-128Hz) and 40Hz stopbands with 0.5\% ripples on bicarbonate (-717Hz), pyruvate (-395Hz) and alanine (-210Hz). The 3D stack-of-spiral trajectory consists of 16 stacks and each stack consists of four 3.8ms interleaves. All gradients have zero net area over the course of one repetition. A 6 pulse non-linear ramp preparation scheme (i.e. 4$^{o}$, 16$^{o}$, 24$^{o}$, 36$^{o}$, 48$^{o}$, 60$^{o}$ for a flip angle of 60$^{o}$) was used to achieve a stable frequency response while the reverse-ordered pulses were used for tip back. The MS-3DSSFP sequence was implemented on a GE Signa MR 3T scanner (GE Healthcare, Waukesha, WI) using a commercial software (RTHawk, HeartVista, Los Altos, CA).

In the bSSFP sequence, TR determines the frequency locations of banding artifacts. A TR of 15.3ms was used for the MS-3DSSFP sequence to maximize the distance between banding artifacts and metabolite frequencies. The excitation profiles of the RF pulse and its averaged transverse magnetization over all echoes of bSSFP acquisitions were simulated. Simulation parameters are: number of RF pulses = 50, TR = 15.3ms, T1 = 30s, T2 = 1s, 6 non-linear ramp preparation pulses, flip angle = 60$^{o}$.

The choice of flip angle for the bSSFP sequence in the hyperpolarized study is a tradeoff between banding artifacts and preserving magnetization for dynamic imaging. Prior bSSFP work \cite{reed2014a} has shown a favorable use of large flip angle ($>$100$^{o}$) to reduce banding artifacts. However, to perform dynamic imaging in hyperpolarized studies, a small flip angle around 30$^{o}$ in MS-GRE acquisitions \cite{nagashima2008a} (equivalent to 60$^{o}$ in the bSSFP sequence) was required to maintain sufficient SNR for multiple time points. In our work, we used a flip angle of 60$^{o}$ to achieve a compromise between the two considerations.

\subsubsection*{Phantom and Animal Experiments}
To test the MS-3DSSFP sequence, phantom experiments were performed on a $^{13}$C-enriched sodium bicarbonate syringe phantom (T1 $\approx$ 26s, T2 $\approx$ 1.5s) with a dual-tuned $^{1}$H/$^{13}$C transceiver birdcage coil. 3D images of the phantom were acquired along with proton images and field maps. To test the excitation profile, $^{13}$C images were acquired with a center frequency offset, both at excitation and at acquisition, by 0Hz, 128Hz, 210Hz, 395Hz, and 717 Hz relative to the phantom frequency, to mimic the images of lactate, pyruvate hydrate, alanine, pyruvate and bicarbonate, respectively. At the frequency of each metabolite, $^{13}$C images were also acquired with small frequency offsets from -30 to 30 Hz with a step of 10Hz. Acquired data were always demodulated to the phantom frequency so that reconstructed images wouldn't be blurred due to off-resonance reconstruction.

Hyperpolarized [1-$^{13}$C]pyruvate animal experiments were performed on healthy Sprague-Dawley rats (N = 3) and transgenic adenocarcinoma of mouse prostate (TRAMP) mice (N = 3) to test our MS-3DSSFP sequence in vivo. $^{13}$C/$^{1}$H birdcage coils (8cm diameter for rats, 5cm diameter for mice) were used. All animal studies were conducted under protocols approved by the University of California San Francisco Institutional Animal Care and Use Committee (IACUC). Both rats and mice were anesthetized with isoflurane (1-2\%) delivered via oxygen gas at 1L/min and placed in a supine position on a heated pad throughout the duration of the experiments. [1-$^{13}$C]pyruvic acid (Sigma Aldrich, St. Louis, MO) mixed with 15mM trityl radical (GE Healthcare, Waukesha, WI) and 1.5mM Gd-DOTA (Guerbet, Roissy, France) was polarized in a 3.35T SPINlab polarizer (GE Healthcare, Waukesha, WI) at 0.8K for $\sim$1h, resulting in a 80mM [1-$^{13}$C]pyruvate solution, with final pH of 6-8. The hyperpolarized [1-$^{13}$C]pyruvate was injected into the animal via tail vein catheters, $\sim$3mL for each rat and $\sim$350$\mu$L for each mouse. 

Hyperpolarized $^{13}$C sequence parameters for animal experiments are shown in Table \ref{tab:protocol}. Each animal received two identical injections of same dose of [1-$^{13}$C]pyruvate. Lactate signals were acquired using the MS-3DSSFP sequence in one injection ("experiment A") but using a 3D MS-GRE sequence (described below) in the other injection ("experiment B"). Pyruvate and alanine signals were acquired using the same 3D MS-GRE sequence in both injections. Such experiment design that different acquisitions were used for lactate while same MS-GRE acquisitions were used for pyruvate and alanine, allows comparing the MS-3DSSFP sequence with a MS-GRE sequence for lactate imaging, as well as examining the perturbation of the MS-3DSSFP sequence on pyruvate and alanine signals. Bicarbonate signals were not acquired due to its inherently low signals and would not provide sufficient signals to compare between the two sequences. Flip angles shown in Table \ref{tab:protocol} were chosen based on previous studies\cite{gordon2019a}. A flip angle of 20$^{o}$ was large enough to see pyruvate signals at a reasonable spatial resolution. A flip angle of 30$^{o}$ for lactate and alanine imaging aimed to maintain sufficient SNR for multiple time points. From preclinical results, we found alanine signals were generally low in the kidneys. Therefore, we chose to use a flip angle of 90$^{o}$ in clinical studies to make sure we can see alanine signals at first couple of time points, although it was not good for acquiring signals over multiple time points.

The 3D MS-GRE sequence consists of a single-band spectral-spatial excitation (130Hz FWHM passband, 870Hz stopband) \cite{gordon2018a} and stack-of-spiral readouts. Each stack was a 22ms single-shot spiral readout. The two injections shared the same spatial resolution, temporal resolution and number of time points. The 3D MS-GRE sequence used 16 excitations for a 3D encoding and each excitation pulse used a flip angle of $7.67^{o}$ so that the equivalent flip angle of these 16 excitations was the same as a $60^{o}$ flip angle used in the MS-3DSSFP sequence. The effective flip angle $\theta_{eq}$ of N excitations with a flip angle of $\theta$ for each excitation is calculated as $\arccos(\cos(\theta)^N)$. Initial pre-scan frequency and power calibration were performed on a $^{13}$C urea phantom which was removed before pyruvate injection. All acquisitions were started 6s after the end of pyruvate injection. For each experiment, a $^{13}$C frequency spectrum was acquired and real-time $^{13}$C B1 calibration \cite{tang2019a} was performed right before metabolite acquisition. 

For rat experiments, an anatomical localizer was acquired using proton 3D bSSFP sequence (FOV 16 $\times$ 16 $\times$ 17.92cm, Matrix size 256 $\times$ 256 $\times$ 112). For the TRAMP mice experiment, an anatomical localizer was acquired using proton T2-weighted fast spin echo sequence (FOV 6 $\times$ 6cm, Matrix size 512 $\times$ 512). For all animal experiments, a B0 map was acquired using IDEAL IQ sequence (FOV 32 $\times$ 32cm, Matrix size 256 $\times$ 256).

\subsubsection*{Human study}
Hyperpolarized [1-$^{13}$C]pyruvate human studies (N = 2) were performed to demonstrate the feasibility of applying the MS-3DSSFP sequence (Figure \ref{fig:seq_scheme}) in the clinical setting. Patients with renal tumors that required surgical removal were recruited under a UCSF institutional review board approved protocol and provided with written informed consent for participation in the study. An Investigational New Drug approval was obtained from the U.S. Food and Drug Administration for generating the agent and implementing the clinical protocol. 1.47g of Good Manufacturing Practices (GMP) [1-$^{13}$C]pyruvate (Sigma Aldrich, St. Louis, MO) mixed with 15mM electron paramagnetic agent (EPA) (AH111501, GE Healthcare, Oslo, Norway) was polarized using a 5T SPINlab polarizer (General Electric, Niskayuna, NY) before being rapidly dissolved with 130$^o$C water and forced through a filter that removed EPA. The solution was then collected in a receiver vessel and neutralized with NaOH and Tris buffer. The receive assembly that accommodates quality-control processes provided rapid measurements of pH, pyruvate and EPA concentrations, polarization, and temperature. In parallel, the hyperpolarized solution was pulled into a syringe (Medrad Inc, Warrendale, PA) through a 0.2$\mu$m sterile filter (ZenPure, Manassas, VA) and transported into the scanner for injection. The integrity of this filter was tested in agreement with manufacturer specifications prior to injection. A 0.43mL/kg dose of $\sim$250mM pyruvate was injected at a rate of 5mL/s via an intra-venous catheter placed in the antecubital vein, followed by a 20mL saline flush. 

In human studies, $^{13}$C kidney images were acquired with in-house built clamshell transmit coil and 8-channel paddle receive array \cite{tropp2011a}. $^{13}$C sequence parameters for this study are presented in Table \ref{tab:protocol}. Similar as the experiment design in animal experiments, the patient received two injections to compare the MS-3DSSFP sequence with MS-GRE sequences. The MS-GRE sequence used in this study was a multi-slice 2D MS-GRE sequence with the same excitation pulse and the same single-shot spiral readout as used in the animal studies. $^{13}$C dynamic imaging was started 6s after the bolus arrival in kidney which was monitored by a bolus tracking sequence \cite{tang2019a}. Initial pre-scan frequency and power calibration were performed on $^{13}$C urea phantom attached outside the receive coil, which was removed before pyruvate injection. Real-time $^{13}$C frequency and power calibration \cite{tang2019a} were performed on the renal tumor and triggered upon bolus arrival. Proton anatomical reference was acquired with a 4 channel paddle receive coil, using a 2D SSFSE sequence with FOV 38 $\times$ 38cm, matrix size 512 $\times$ 512.

\subsubsection*{Reconstruction and Data Analysis}
For all studies, gridding of k-space data were performed using Kaiser-Bessel gridding method \cite{jackson1991a} (http://web.stanford.edu/class/ee369c/mfiles/gridkb.m) with an oversampling factor of 1.4 and a kernel width of 4.5. The gridded k-space data was then inverse Fourier transformed to the reconstructed image. Multi-channel data were combined by using pyruvate signals as coil sensitivity maps \cite{zhu2019a}. For display purposes, images were zero-filled by a factor of 2 and applied with a 2D fermi filter.

Area-under-the-curve (AUC) images were calculated by summing the complex data through time. Signal-to-noise ratio (SNR) was calculated as signal magnitude divided by the standard deviation of the real part of the noise. Lactate-to-pyruvate AUC ratio images were calculated by dividing the SNR of lactate AUC images by the SNR of pyruvate AUC images. To compare AUC of a metabolite between experiment A (pyruvate and alanine: MS-GRE; lactate: MS-3DSSFP) and experiment B (pyruvate, lactate, and alanine: MS-GRE) (Table \ref{tab:protocol}), SNR of the AUC images was calculated and then divided by the SNR of pyruvate AUC images acquired in the same experiment. To compare dynamic curves of a metabolite between experiment A and experiment B, SNR of each time point was calculated and then divided by the highest SNR of the pyruvate dynamic curve acquired in the same experiment.

Signal levels of undesired metabolites in MS-3DSSFP lactate acquisitions were estimated. First, the concentration ratio between an undesired metabolite and lactate was estimated using the signals acquired from experiment B where all compounds were acquired with MS-GRE sequences. Flip angle was compensated in the concentration ratio. Next, to estimate the signal ratio between an undesired metabolite and lactate in MS-3DSSFP, the concentration ratio was multiplied with MS-3DSSFP point spread function (PSF) amplitude ratio between the undesired metabolite and lactate. The MS-3DSSFP PSF amplitude was calculated by multiplying MS-3DSSFP excitation profile with the simulated PSF amplitude of the MS-3DSSFP readout. The following equation describes the above calculation:
\begin{equation} \label{eq:contribution}
p_x = \frac{S_x * sin(\theta_l) * \delta_x * I_x}{S_l * sin(\theta_x)}
\end{equation}
where $p$ is the signal level (\%) of an undesired metabolite $x$ in MS-3DSSFP lactate acquisitions, $l$ is lactate, $S$ is the signal measured in experiment B, $\theta$ is the flip angle used in experiment B, $\delta$ is the stopband amplitude of the excitation RF pulse used in the MS-3DSSFP sequence, $I$ is the central amplitude of the simulated off-resonance PSF of the interleaved spiral readouts used in the MS-3DSSFP sequence. Pyruvate hydrate signals were assumed to be as 8\% of pyruvate signals \cite{golman2006a}. Stopband amplitudes $\delta$ are described in pulse design: 0.5\% for alanine, 0.5\% for pyruvate and 5\% for pyruvate hydrate. The off-resonance PSF amplitudes $I$ of the MS-3DSSFP readouts are obtained from simulations (Supporting Information Figure S1): 0.327 for alanine, 0.191 for pyruvate and 0.701 for pyruvate hydrate.


\section*{Results}
\begin{figure} 
\centering  
\includegraphics[width=5.5in]{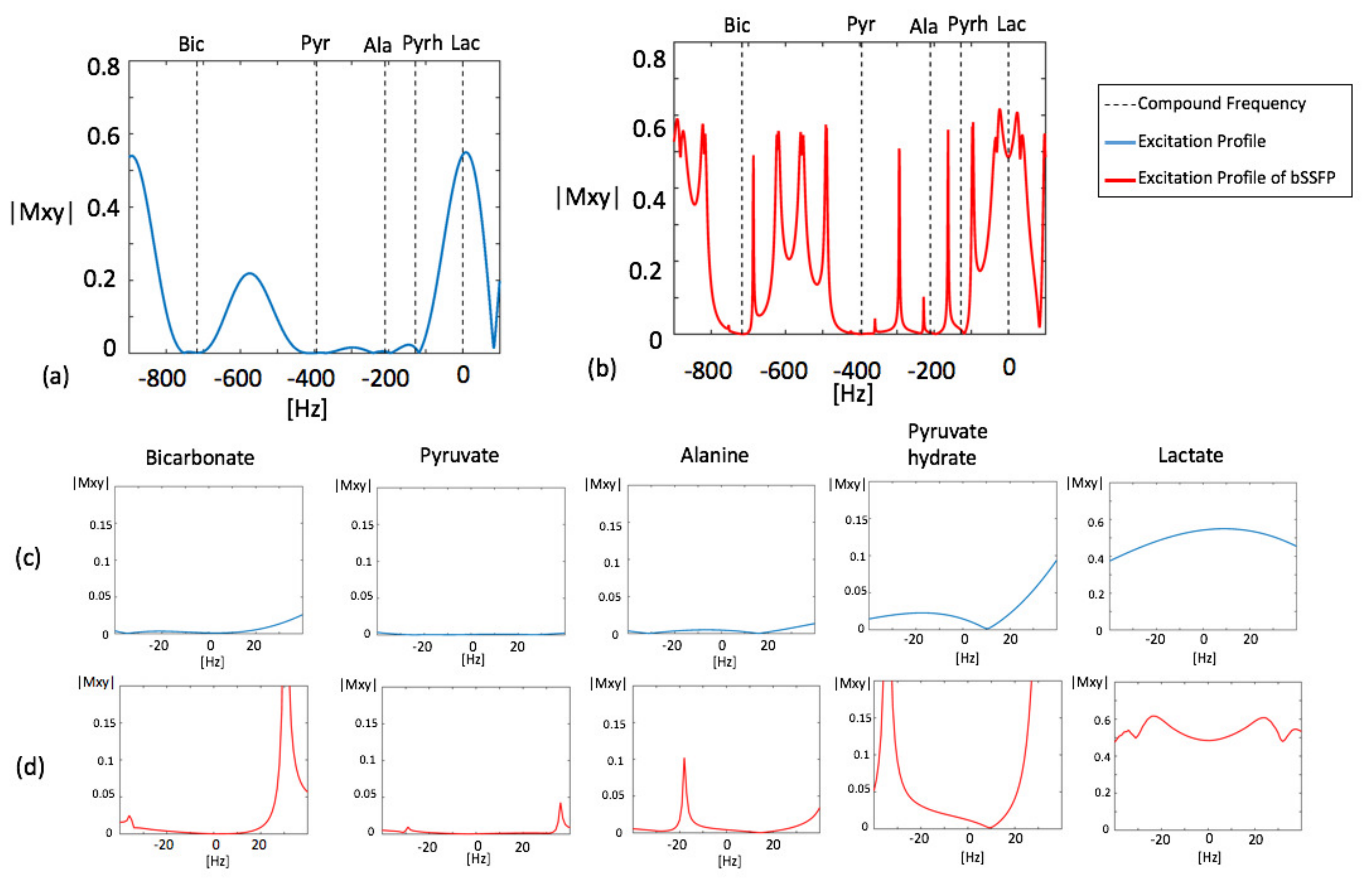}
  \caption
    {
Simulated excitation profiles of the excitation pulse alone (blue) and its averaged transverse magnetization over 64 pulses of a bSSFP acquisition (red). An overall view of the profile is shown in graph (a) and graph (b). Zoomed views ($\pm$40Hz) of excitation profiles around each metabolite are shown in graph (c) and graph (d). The excitation pulse has a 40Hz passband on lactate (0Hz), a 40Hz stopband of 5\% maximum ripple on pyruvate hydrate (-128Hz) and 40Hz stopbands of 0.5\% maximum ripples on bicarbonate (-717Hz), pyruvate (-395Hz) and alanine (-210Hz). Simulation parameters for bSSFP acquisitions include: number of RF pulses = 64, TR = 15.3ms, T1 = 30s, T2 = 1s, 6 non-linear ramp preparation pulses, flip angle = 60$^{o}$
    }
  \label{fig:excitation_profile}
\end{figure}

Simulated excitation profiles of the MS-3DSSFP sequence and its averaged transverse magnetization over all bSSFP echoes are shown in Figure \ref{fig:excitation_profile}. Frequency bands and stopband ripples of the excitation profiles were as desired. Most banding artifacts fell outside of the desired frequency bands except one banding artifact which was observed 18Hz upfield from the alanine frequency. In the simulation, the amplitude of this banding artifact was about a third of the on-resonance peak, although its actual value in a hyperpolarized $^{13}$C pyruvate study depends on the T1 and T2 of alanine as well as the conversion rate between pyruvate and alanine.

\begin{figure} 
\centering  
  \includegraphics[width=5.5in]{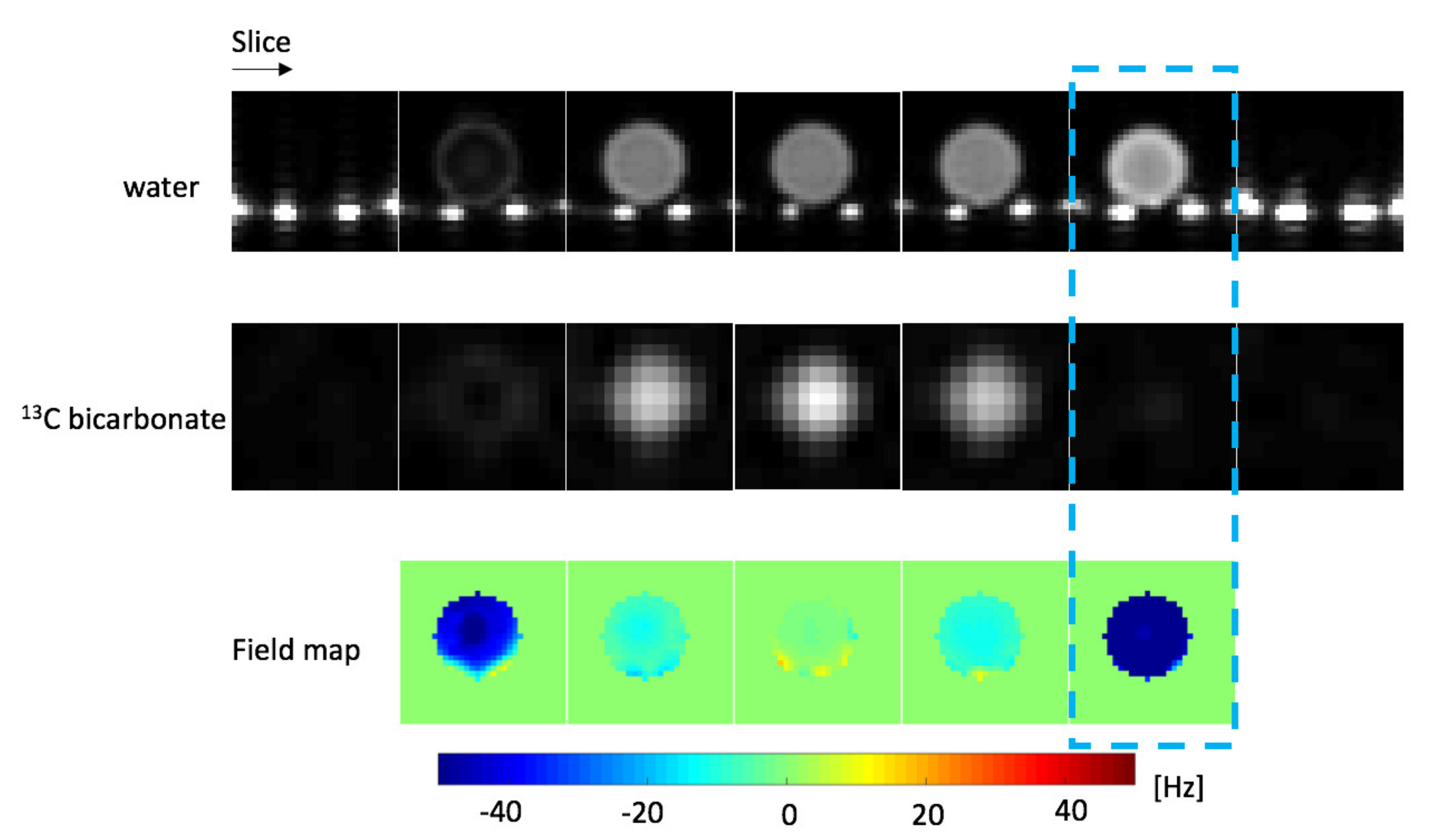}
  \caption
    {
Validation of the MS-3DSSFP sequence on a [$^{13}$C]bicarbonate syringe phantom (T1 $\sim$= 26s, T2 $\sim$= 1.5s) with a rat birdcage coil. $^{13}$C images were acquired at 8 $\times$ 8 $\times$ 20mm and reconstructed at 4 $\times$ 4 $\times$ 20mm.  Bright spots at the bottom of proton images are the water pad in the coil. B0 maps are shown at $^{13}$C frequency. The dash-line-boxed slice shows a bright proton image but a dark $^{13}$C image, consistent with the large B0 variation (-50Hz) at that slice.
    }
  \label{fig:phantom}
\end{figure}

Results of validating the MS-3DSSFP sequence on a [$^{13}$C]bicarbonate syringe phantom (T1 $\sim$= 26s, T2 $\sim$= 1.5s) with a rat birdcage coil are shown in Figure \ref{fig:phantom}. $^{13}$C images, proton images and B0 maps scaled to $^{13}C$ frequency were provided. The dash-line-boxed slice shows a bright proton image but a dark $^{13}$C image. This is consistent with the large B0 variation (-50Hz) at that slice. Results of validating the excitation profile of the bSSFP sequence on the phantom are presented in Supporting Information Figure S2. Excitation profiles measured from phantom experiments were found to be consistent with the simulation.

\begin{figure}  
\centering  
\includegraphics[width=6.5in]{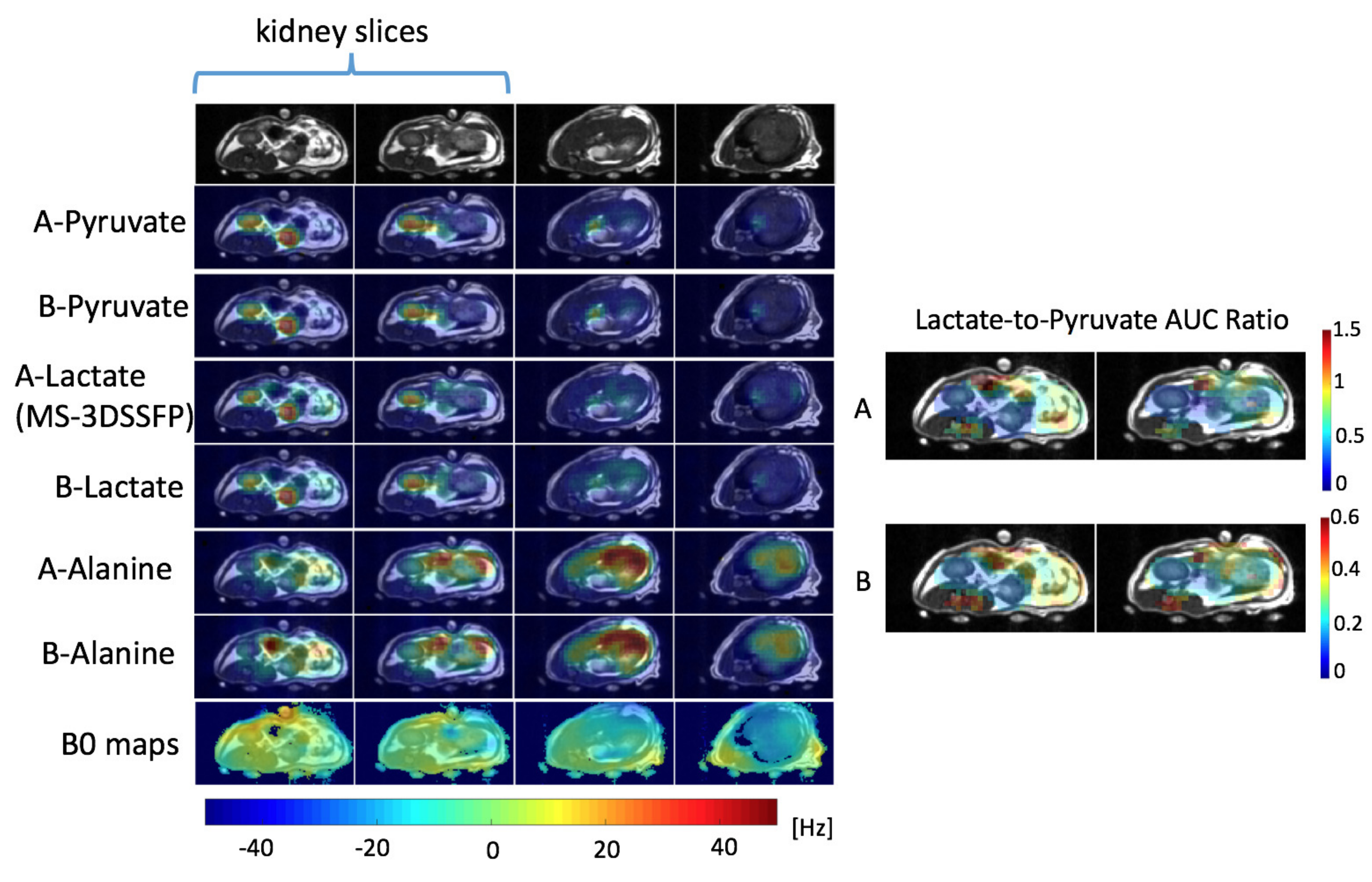}
  \caption
    {
Comparison of the MS-3DSSFP sequence with a 3D MS-GRE sequence on a healthy rat with hyperpolarized [1-$^{13}$C]pyruvate injections using experiment A (pyruvate and alanine: MS-GRE; lactate: MS-3DSSFP) and experiment B (pyruvate, lactate and alanine: MS-GRE). Experiment parameters are described in Table \ref{tab:protocol}. AUC maps of each metabolite are displayed at the maximum signal across slices. B0 maps were thresholded using a mask removing the pixels with SNR lower than 3 in the water magnitude images. B0 maps colorbar is displayed at the bottom. Lactate-to-pyruvate AUC ratio maps of the kidney slices are shown and thresholded using a mask removing the pixels with SNR lower than 3 in the lactate or pyruvate magnitude image.
    }
  \label{fig:rat}
\end{figure}

Results of representative hyperpolarized [1-$^{13}$C]pyruvate experiments to compare the MS-3DSSFP sequence with MS-GRE sequences using the experiment parameters in Table \ref{tab:protocol} are presented in Figure \ref{fig:rat} (rat), Figure \ref{fig:tramp} (TRAMP mouse) and Figure \ref{fig:human} (renal tumor patient). Comparing lactate AUC maps of the two experiments, no banding artifacts were observed in the MS-3DSSFP results. This finding agrees with the homogenous B0 maps found in most areas of rat kidneys, TRAMP tumors and human kidneys, although large B0 variations are found near the tissue-air interface. In some tumor regions, lactate-to-pyruvate AUC ratio maps reveal different contrasts between the two experiments, as shown in Figure \ref{fig:tramp} and Figure \ref{fig:human}. Compared to results of experiment B (pyruvate: MS-GRE; lactate: MS-GRE), lactate-to-pyruvate AUC ratio map of experiment A (pyruvate: MS-GRE; lactate: MS-3DSSFP) shows better alignment with the underlying T2 weighted proton images. This could be a result of T2 contrast provided by the MS-3DSSFP sequence. Higher values of lactate-to-pyruvate AUC ratio map are found in experiment A compared to experiment B, demonstrating the MS-3DSSFP sequence provides higher SNR over MS-GRE sequences. AUC maps of pyruvate and alanine show consistent contrast between the two experiments, demonstrating minimal perturbation of the newly designed RF pulse on pyruvate and alanine.

Representative dynamic curves of lactate, pyruvate and alanine signals are presented in Figure \ref{fig:dynamic} acquired with experiment parameters described in Table \ref{tab:protocol}. Metabolites signal ratios between the two experiments are presented in Figure \ref{fig:metabolite_auc}. Compared with MS-GRE sequences, the MS-3DSSFP sequence shows an overall approximately 2.5X SNR improvement and demonstrates higher SNR performance at every time point for lactate imaging in rat kidneys, tumors of TRAMP mice and human kidneys. Comparing AUC between the two experiments, there is almost no difference in pyruvate and a 5\% to 20\% difference in alanine AUC, which demonstrates the lactate spectral selectivity of the MS-3DSSFP sequence.

Signal levels of undesired metabolites (i.e. pyruvate and alanine) in MS-3DSSFP lactate acquisitions were quantified in three types of ROIs: rat kidneys, tumors of TRAMP mice and human kidneys (Table \ref{tab:contamination}). The highest signal contribution from undesired metabolites was found in rat kidneys - 0.07\% from alanine, 1.24\% from pyruvate and 3.64\% from pyruvate hydrate, where 1\% means that the ratio between the undesired metabolite and lactate is 1\%. 

\section*{Discussion}


\subsubsection*{Metabolite Specific Excitation for bSSFP}
We designed a multiband RF pulse (Figure \ref{fig:seq_scheme}) under the constraint of short TR in a bSSFP sequence to achieve spectrally selective excitation on lactate while minimally perturbing other metabolites on a clinical 3T scanner. The 9ms pulse duration was primarily determined by the frequency difference between lactate and pyruvate hydrate (128Hz at 3T) which has the closest frequency to lactate among all compounds in hyperpolarized [1-$^{13}$C]pyruvate studies. The newly designed RF pulse had a maximum power of 0.2195G and did not experience specific absorption rates (SAR) issues in our studies.

Our MS-3DSSFP sequence can be easily adapted to image [1-$^{13}$C]pyruvate or [1-$^{13}$C]bicarbonate. Because these two metabolites have larger frequency differences from other compounds compared to the frequency difference of lactate to pyruvate hydrate, it is guaranteed to find a solution of metabolite specific excitation pulse for these two metabolites while meeting the TR requirement in our studies. In contrast, it is challenging to design a metabolite specific excitation pulse for imaging alanine, which has a frequency difference of 82Hz (at 3T) to pyruvate hydrate, much closer than the frequency difference between lactate to pyruvate hydrate.

\subsubsection*{Spiral bSSFP vs Cartessian bSSFP}
Spiral readouts were used in the MS-3DSSFP to accelerate the acquisition. All prior HP $^{13}$C bSSFP work used Cartesian readouts, which brought challenges to acquire multiple metabolites in dynamic imaging. For example, whole human brain HP $^{13}$C imaging typically uses a matrix size of 16 $\times$ 16 $\times$ 16 and a FOV of 24 $\times$ 24 $\times$ 24 cm. Assuming a TR of 15ms, 3D Cartesian readouts need 16 $\times$ 16 $\times$ 15ms = 3.84s to cover a volume for one metabolite, which would result in insufficient temporal resolution when more than one metabolite needs to be acquired. Given the relatively small matrix size,  undersampling strategies will only achieve limited acceleration, therefore fast imaging readouts are preferred\cite{durst2015a}. Under the same requirement of matrix size and FOV, stack-of-spiral readouts using two interleaves per stack could achieve an acquisition time of 2 $\times$ 16 $\times$ 15ms = 0.48ms and an even larger matrix size (24 x 24 x 24), assuming a 3.8ms readout time for each spiral interleaf (same as what we used in this study), a 5 G/cm maximum gradient and a 20 G/cm/ms maximum slew rate.

The center k-space of the spiral readout is not at the center of the TR, which may cause a slight SNR loss compared to Cartesian readouts. Assuming a spiral readout duration of 3.8 ms as used in our studies and a T2$^{*}$ of 50 ms, SNR loss of using spiral readouts will be 1-exp(-1.9/50) = 4\%.

\subsubsection*{MS-3DSSFP vs MS-GRE}
By comparing the results of two experiments whose experiment parameters are shown in Table \ref{tab:protocol} (experiment A: MS-3DSSFP for lactate, MS-GRE for pyruvate and alanine; experiment B: MS-GRE for lactate, pyruvate and alanine.), we assess the performance of the MS-3DSSFP sequence in the aspects of SNR, contrast, banding artifacts, artifacts by exciting undesired metabolites and impact on acquisition of other metabolites. These issues will be discussed in the following paragraphs.

To fairly compare SNR between the MS-3DSSFP sequence and MS-GRE sequences, we used the same spatial resolution, starting time of acquisition, temporal resolution and number of time points. The effective flip angle (see definition in Methods) of the MS-GRE sequence was the same as the flip angle used in the MS-3DSSFP sequence. Readout durations of the two sequences were not matched since T2$^{*}$ limits the readout duration of a MS-GRE sequence. Compared with MS-GRE sequences, our MS-3DSSFP sequence has shown an overall 2.5-fold SNR improvement (Figure \ref{fig:dynamic} and Figure \ref{fig:metabolite_auc}) for dynamic lactate imaging in hyperpolarized [1-$^{13}$C]pyruvate studies. The SNR improvement ratio would increase with T2 (Supporting Information Figure S3). Besides utilizing T2, other features of the MS-3DSSFP sequence could also contribute to the SNR improvement, including shorter echo time due to shorter RF pulse and shorter spiral readout time which results in less signal reduction caused by B0 inhomogeneity. 

\begin{figure} 
\centering  
  \includegraphics[width=5.5in]{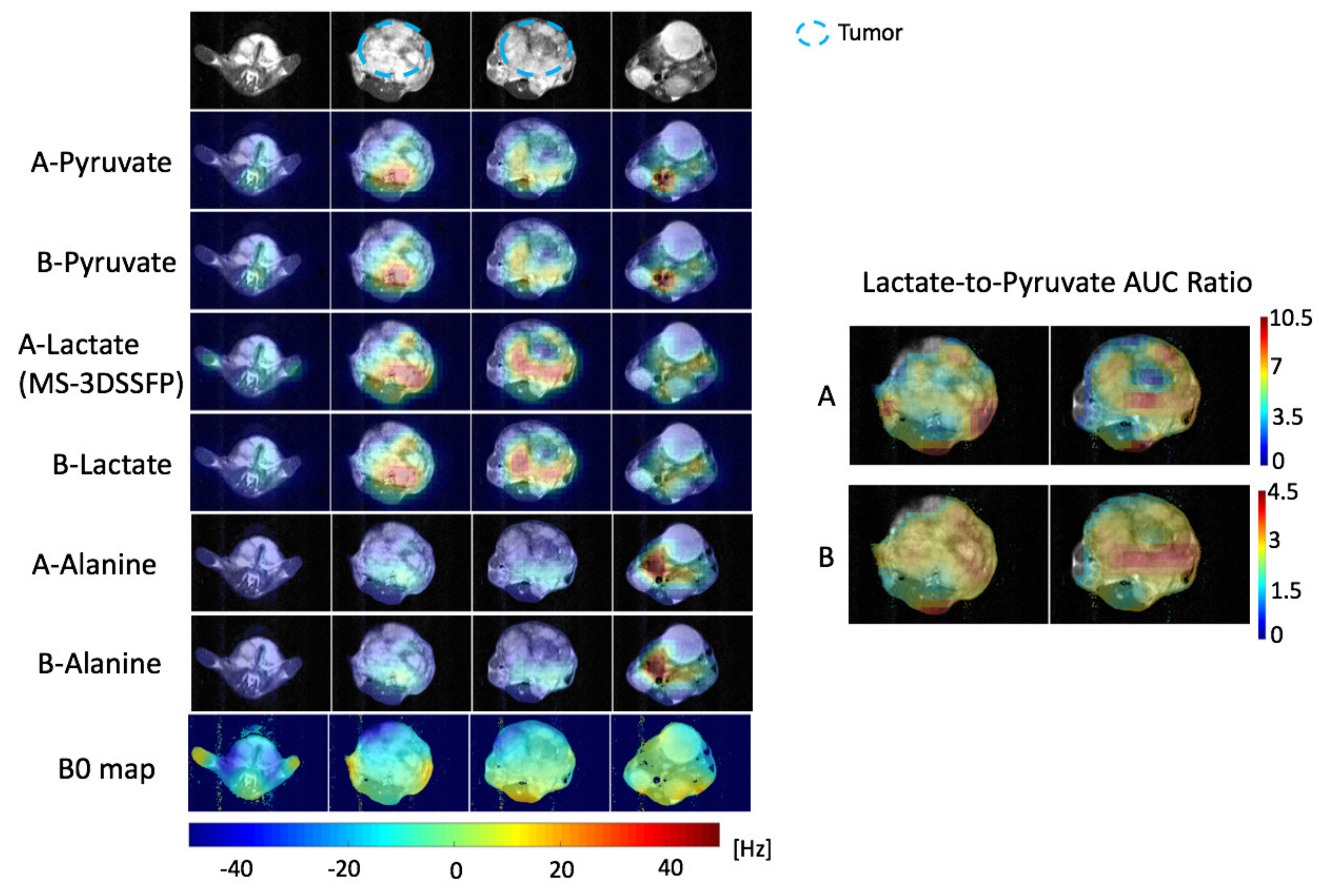}
  \caption
    {
Comparison of the MS-3DSSFP sequence with a 3D MS-GRE sequence on a TRAMP mouse prostate tumor with hyperpolarized [1-$^{13}$C]pyruvate injections using experiment A (pyruvate and alanine: MS-GRE; lactate: MS-3DSSFP) and experiment B (pyruvate, lactate and alanine: MS-GRE). Experiment parameters are described in Table \ref{tab:protocol}. AUC images of each metabolite, and lactate-to-pyruvate AUC ratio images and B0 maps are shown.  B0 and AUC ratio maps are thresholded the same way as described in Figure \ref{fig:rat}. B0 maps colorbar is displayed at the bottom.
    }
  \label{fig:tramp}
\end{figure}

\begin{figure}  
\centering  
  \includegraphics[width=1\textwidth]{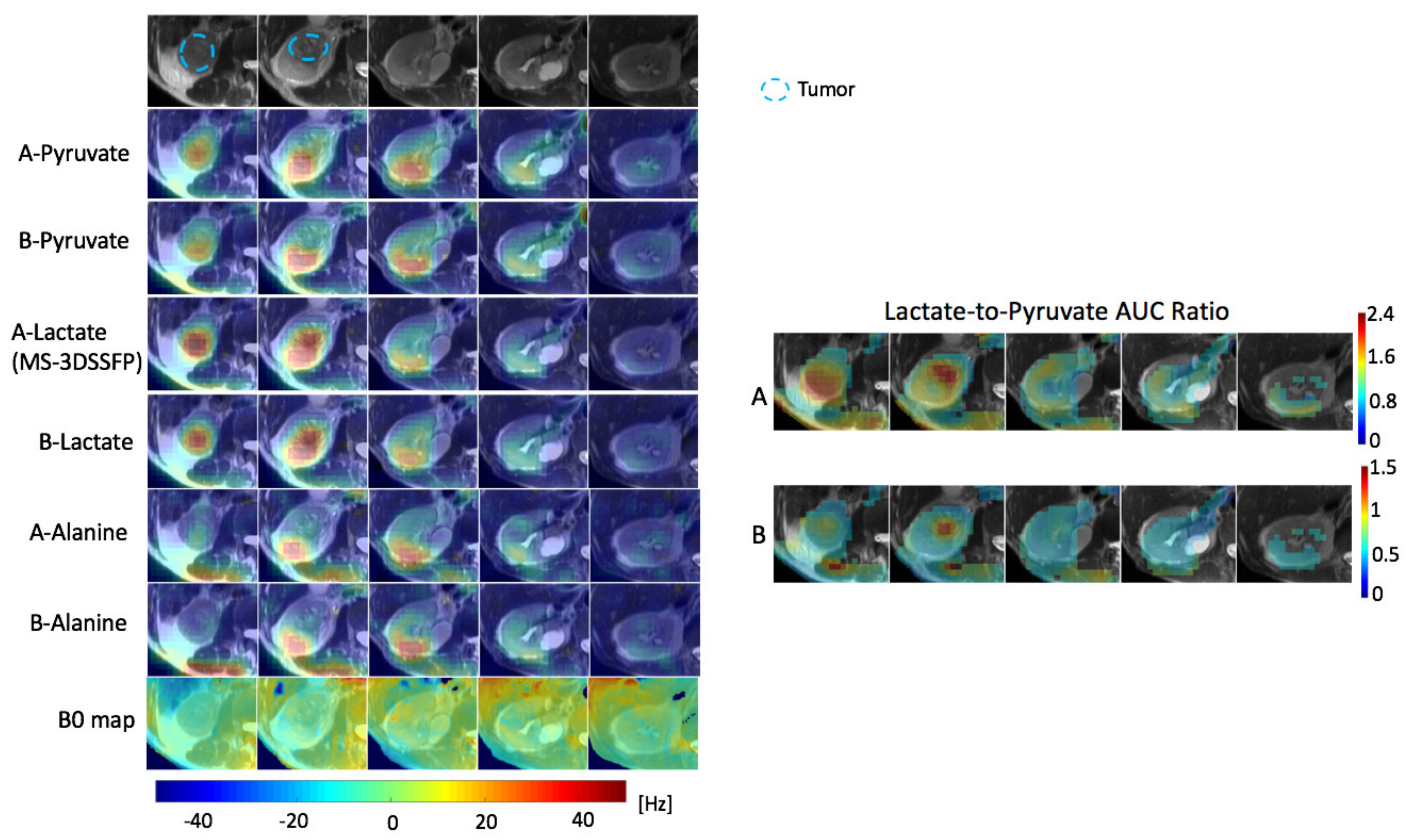}
   
  \caption
    {
Comparison of the MS-3DSSFP sequence with a multi-slice 2D MS-GRE sequence in the kidneys of a patient with a renal tumor with hyperpolarized [1-$^{13}$C]pyruvate injections using experiment A (pyruvate and alanine: MS-GRE; lactate: MS-3DSSFP) and experiment B (pyruvate, lactate and alanine: MS-GRE). Experiment parameters are described in Table \ref{tab:protocol}. AUC images of each metabolite, and lactate-to-pyruvate AUC ratio images and B0 maps are shown. B0 and AUC ratio maps are thresholded the same way as described in Figure \ref{fig:rat}. B0 maps colorbar is displayed at the bottom.
    }
  \label{fig:human}

\end{figure}

\begin{figure}  
\centering  
  \includegraphics[width=1\textwidth]{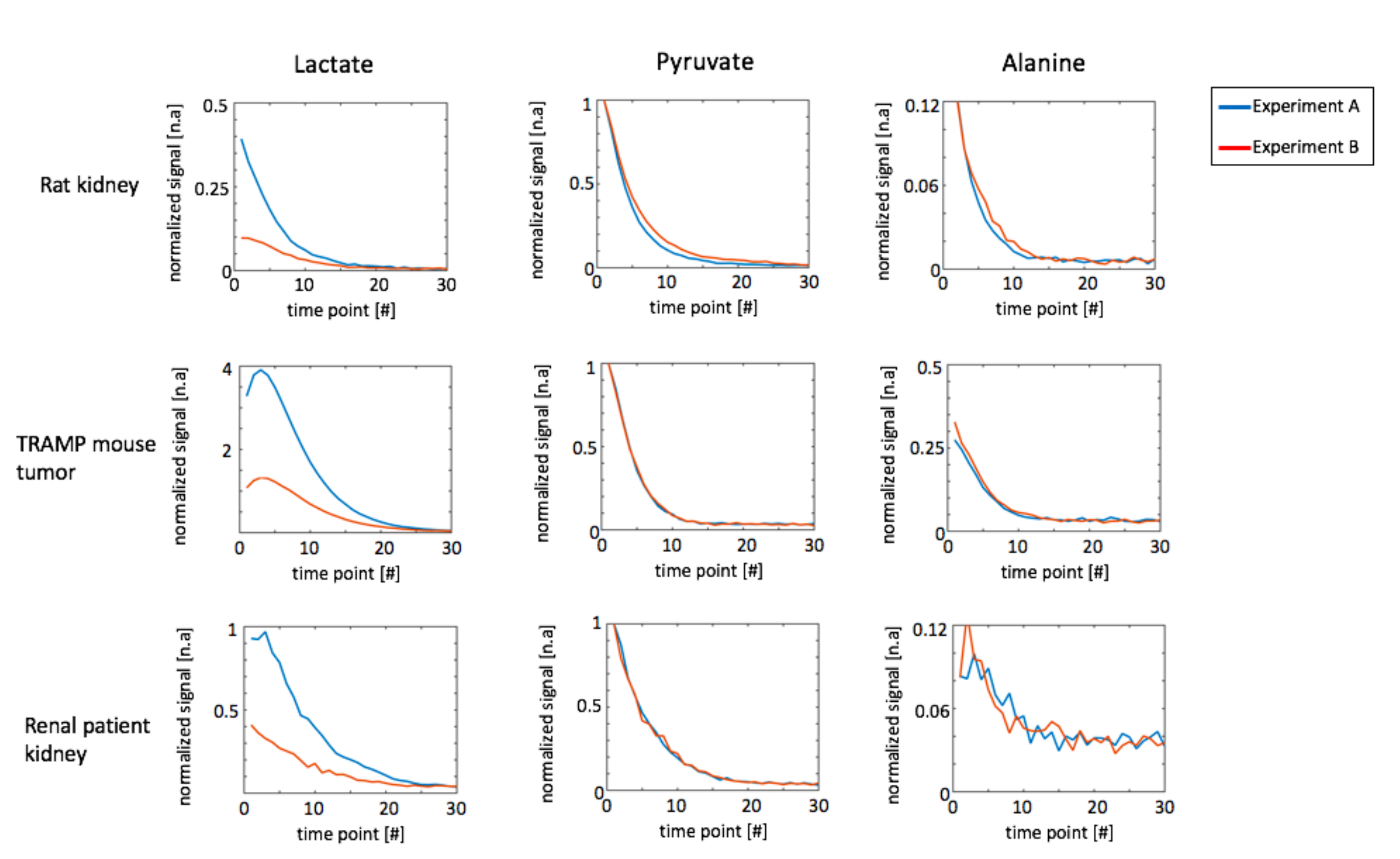}
    \caption
    {
Representative dynamic curves of lactate, pyruvate and alanine signals acquired in experiment A (pyruvate and alanine: MS-GRE; lactate: MS-3DSSFP) and experiment B (pyruvate, lactate and alanine: MS-GRE). Experiment parameters are described in Table \ref{tab:protocol}. All signals were divided by corresponding noise signals and then divided by the highest value of the pyruvate dynamic curve. Corresponding dynamic images are shown in Supporting Information Figure S4, Supporting Information Figure S5 and Supporting Information Figure S6.
    }  
 \label{fig:dynamic}
 
\end{figure}

\begin{figure}  
\centering  
\includegraphics[width=6in]{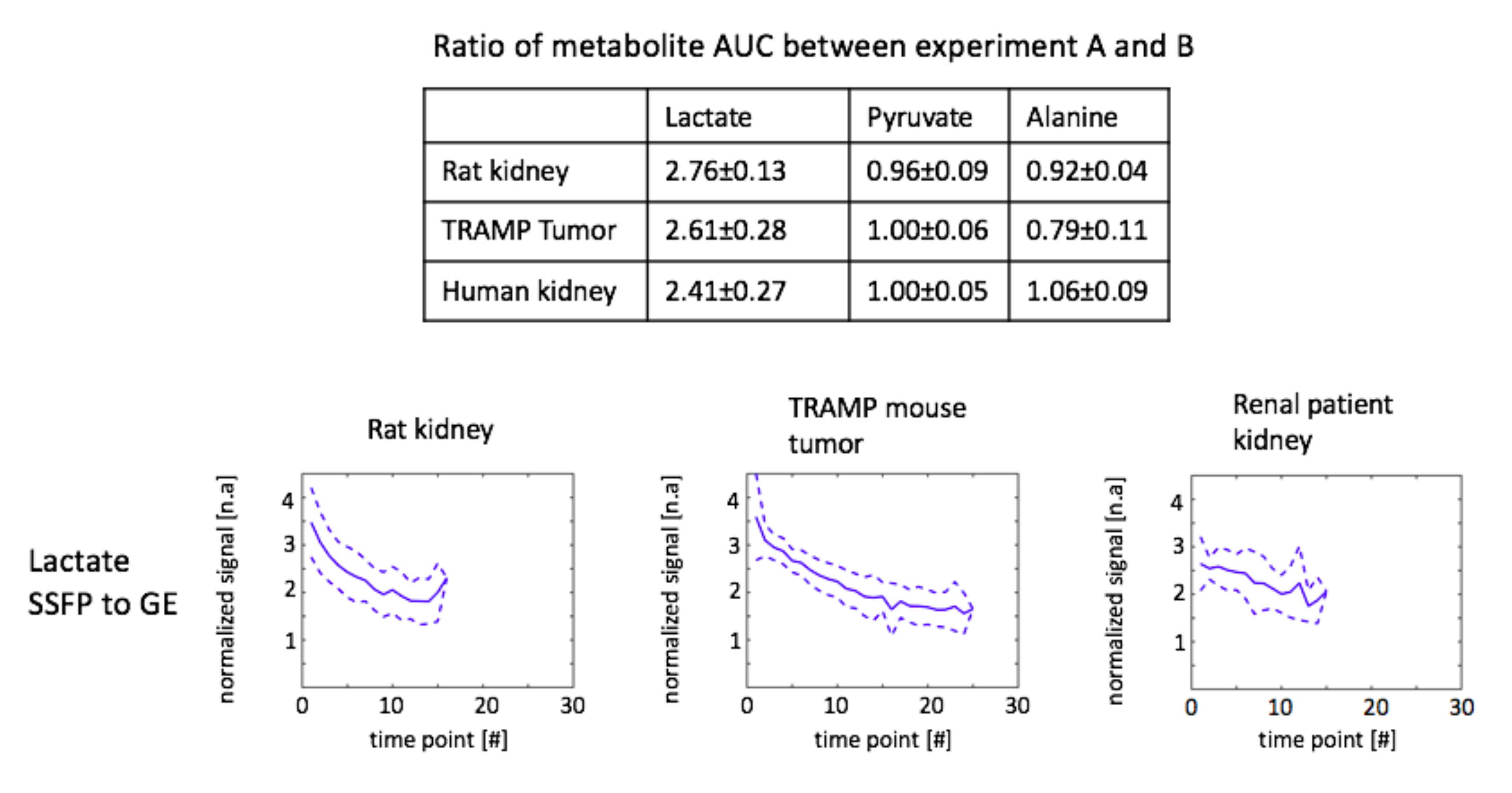}
 \caption
    {
Metabolites AUC ratios and lactate ratios of dynamic curves between experiment A (pyruvate and alanine: MS-GRE; lactate: MS-3DSSFP) and experiment B (pyruvate, lactate and alanine: MS-GRE) at each time point. Experiment parameters are described in Table \ref{tab:protocol}. Data of rat kidneys, TRAMP tumors and human kidneys were included in the summary with a criterion of SNR greater than 3. The averaged lactate ratios are shown by the solid lines and $\pm$1 standard deviations are shown by the dashed lines.
    }
  \label{fig:metabolite_auc}

\end{figure}

Compared with MS-GRE sequences, the MS-3DSSFP sequence can also provide T2 contrast for tissue characterization. In our studies, such contrast differences are potentially observed in some tumor regions in lactate-to-pyruvate AUC ratio maps as shown in Figure \ref{fig:tramp} and Figure \ref{fig:human}.Parameters of the MS-3DSSFP sequence (e.g. flip angle, TR) could be explored to enable jointly estimating lactate T2 and pyruvate-to-lactate conversion rate.

Excitation profiles of a bSSFP sequence are determined by both the excitation profile of the RF pulse and banding artifacts governed by the chosen TR. Two types of image artifacts could be a result of excitation profiles of a bSSFP sequence: null-signal banding artifacts of metabolites of interest and artifacts by exciting undesired metabolites. Comparing in vivo lactate AUC results (Figure \ref{fig:rat}, Figure \ref{fig:tramp}, Figure \ref{fig:human}) of MS-GRE and MS-3DSSFP, no null-signal banding artifacts are found in the MS-3DSSFP results. In some peripheral regions of TRAMP tumors (Figure \ref{fig:tramp}) where large B0 variations are noted, both MS-GRE and MS-3DSSFP sequences shows hypointense signals in AUC maps of pyruvate and lactate but show hyperintense signals in pyruvate-to-lactate AUC ratio maps. This indicates that both MS-GRE and MS-3DSSFP sequences are sensitive to B0 inhomogeneity for different reasons.For MS-3DSSFP, the reason is reduced excitation due to a narrow excitation bandwidth (40Hz). For MS-GRE, reduced excitation could also be a reason although not as worse as MS-3DSSFP, and another reason could be reduced signal due to long spiral readouts (22ms), which can be improved by using proper off-resonance correction \cite{noll1992a,man1997a}. 

Exciting undesired metabolites would cause both artificially elevated lactate signals and ring-shaped artifacts. Simulation (Supporting Information Figure S1) shows that for the interleaved spiral readouts used in our studies, artifacts from pyruvate hydrate and alanine mostly stay in the center of the point spread function while artifacts from pyruvate, urea and bicarbonate are spread out, which will cause ring-shaped artifacts. No ring-shaped artifacts are observed in the MS-3DSSFP images. Signal levels of undesired metabolites (i.e. pyruvate and alanine) in MS-3DSSFP lactate acquisitions were summarized in Table \ref{tab:contamination}. These signal levels could be higher if acquisition starts early when pyruvate and pyruvate hydrate signals are high while lactate signals are yet to build up.

Exciting undesired metabolites would also sacrifice their magnetization and reduce their signals. Comparing AUC results between the two experiments (Figure \ref{fig:metabolite_auc}), there is almost no difference in pyruvate and a 5\% to 20\% difference in alanine AUC. The cost in alanine signals is consistent with the simulation (Figure \ref{fig:excitation_profile}) and phantom (Supporting Information Figure S2) results where a banding artifact is identified 18Hz upfield from alanine frequency. Therefore, it is more robust to apply our MS-3DSSFP sequence for imaging ROIs with low alanine productions such as most tumors, kidney or brain.

\subsubsection*{Interleaving Different Sequences In One Injection}
In hyperpolarized $^{13}$C studies, signals of different metabolites are usually acquired using the same sequence so that results reveal the contrast of metabolite concentration. In our studies, we developed a method of imaging different metabolites using different sequences in one injection, i.e., imaging lactate using MS-3DSSFP while imaging pyruvate and alanine using MS-GRE. This method is achieved by using a commercial software (RTHawk, HeartVista, Los Altos, CA) where interleaving different sequences are easily incorporated. It could potentially provide multiple contrasts for multiple metabolites in a single injection, whereas the same purpose could possibly be achieved by using a MR-Fingerprinting type of acquisition\cite{ma2013a}.


\subsubsection*{Precautions of Performing MS-3DSSFP Experiments}
To run the MS-3DSSFP sequence, several issues need to be carefully handled. As discussed before, the RF pulse used in the MS-3DSSFP sequence has a narrow bandwidth (40Hz) and real-time frequency calibration \cite{tang2019a} is crucial to the robustness of this sequence. Furthermore, the multiband RF pulse used in this study does not avoid exciting urea, therefore, it is suggested to remove the urea phantom which was used in pre-scan frequency and power calibration prior to pyruvate injection, otherwise there could be spiral off-resonance artifacts from urea signals. Finally, the multiband RF pulse was not slice-selective, therefore the field of view along the slice direction needs to be as large as the extent of $^{13}$C receive coils.

\section*{Conclusion}
This work describes a novel 3D bSSFP sequence that integrates a lactate specific excitation pulse and stack-of-spiral readouts for improved lactate dynamic imaging in hyperpolarized [1-$^{13}$C]pyruvate studies on a clinical 3T scanner. Compared with MS-GRE sequences, the MS-3DSSFP sequence showed an overall 2.5X SNR improvement for lactate imaging in rat kidneys, tumors of TRAMP mice and human kidneys. Future work will include exploring joint estimation of lactate T2 and pyruvate-to-lactate conversion rate, extending the applications of the proposed sequence for imaging regions with acceptable B0 homogeneity such as human brain, as well as imaging other metabolites (e.g. pyruvate, bicarbonate) in hyperpolarized [1-$^{13}$C]pyruvate studies.

\subsection*{Acknowledgements}
The authors thank Lucas Carvajal, Jennifer Chow, Hsin-yu Chen, Justin Delos Santos, James Slater, Namasvi Jariwala, Mary Mcpolin, Kimberly Okamoto for their help on the project. This work was supported by the National Institute of Biomedical Imaging and Bioengineering (P41EB013598, R01EB016741, U01EB026412), the American Cancer Society (RSG-18-005-01-CCE), and a UCSF Research Evaluation and Allocation Committee Shared Instrument Award.
\clearpage

\listoffigures
\clearpage
\listoftables
\clearpage

%


\begin{table}[h] 
\centering  
\includegraphics[width=6in]{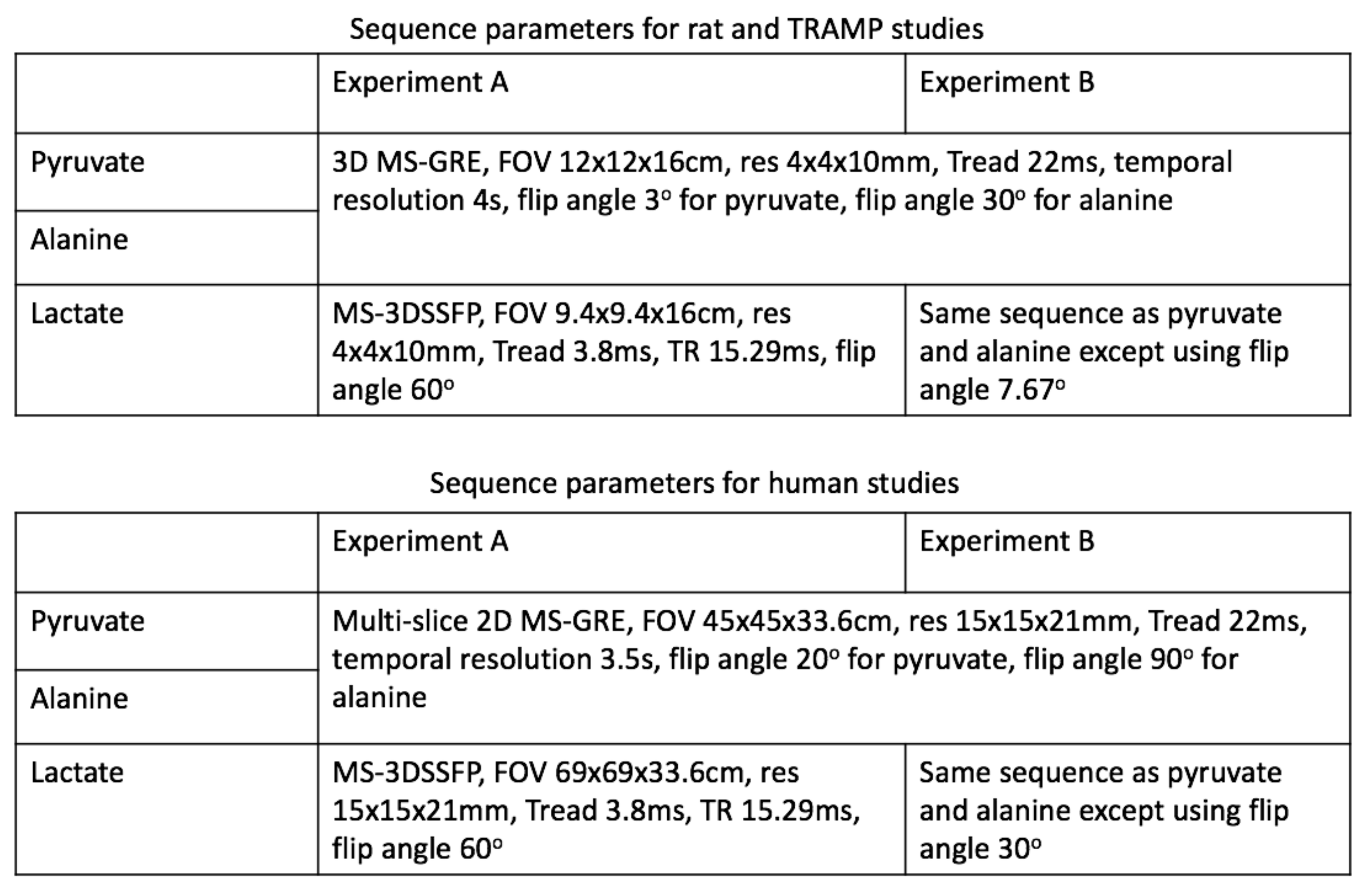}
 \caption
    {
$^{13}$C sequence parameters used in rat, TRAMP and human studies with hyperpolarized [1-$^{13}$C]pyruvate injection. For the same subject, two experiments (A and B) would be performed back-to-back for comparison. In experiment A, lactate signals were acquired with the metabolite specific 3D SSFP (MS-3DSSFP) sequence while pyruvate and alanine signals were acquired with the metabolite specific GRE (MS-GRE) sequences. In experiment B, all three metabolites were acquired with MS-GRE sequences. In TRAMP mouse studies and one of the human study, experiment B was performed first. In other studies, experiment A was performed first.
    }
  \label{tab:protocol}
\end{table}

\begin{table}[h]  
\centering  
\includegraphics[width=6in]{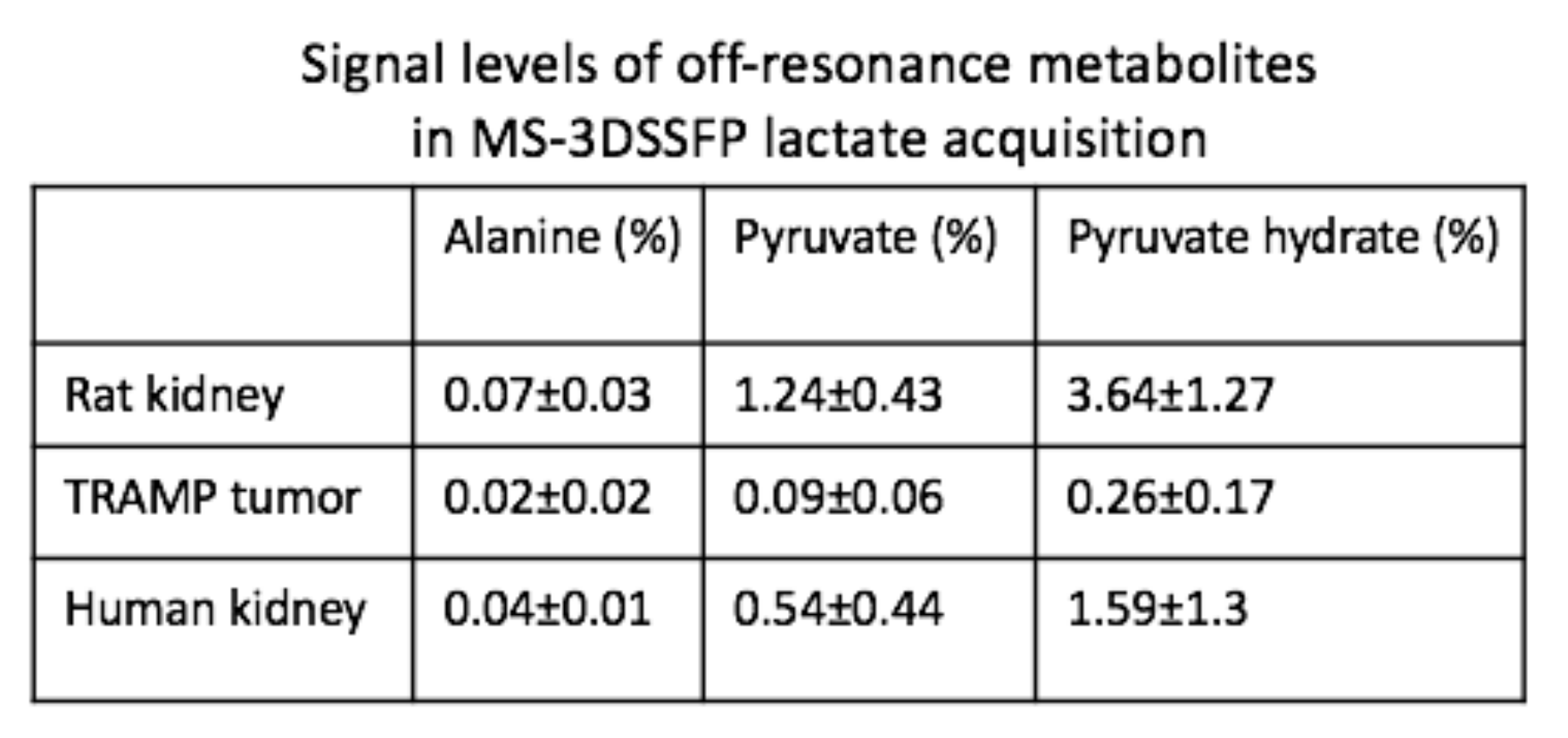}
 \caption
    {
Estimated signal levels of off-resonance metabolites in lactate acquisitions using the MS-3DSSFP sequence. Signal levels are estimated according to Eq. 1. A signal level of 1\% means that in a MS-3DSSFP lactate acquisition, the ratio between the off-resonance metabolite and lactate is 1\%. 
    }
  \label{tab:contamination}
\end{table}

\newpage
\section*{Supporting Information}

\begin{enumerate}[label=S\arabic*]
\item{Simulations of off-resonance PSF of the interleaved spiral readouts (4 interleaves, 3.8ms for each interleaf) used in this study. Frequencies of metabolites are Lactate = 0Hz, Pyruvate Hydrate = -128Hz, Alanine = -210Hz, Pyruvate = -395Hz, Urea = -635Hz, Bicarbonate = -717Hz.}
\label{figS:sim_offres}

\item{Validation of the MS-3DSSFP sequence on a $^{13}$C-enriched sodium bicarbonate phantom (T1 ~= 26s, T2 ~= 1.5s). $^{13}$C images were acquired with a center frequency offset, both at excitation and at acquisition, by 0Hz, 128Hz, 210Hz, 395Hz, and 717 Hz relative to the phantom frequency, to mimic the images of lactate ("Lac"), pyruvate hydrate ("Pyrh"), alanine ("Ala"), pyruvate ("Pyr") and bicarbonate ("Bic"), respectively. At the frequency of each metabolite, $^{13}$C images were also acquired with small frequency offsets from -30 to 30 Hz with a step of 10Hz. Acquired data were always demodulated to the phantom frequency so that reconstructed images wouldn't be affected by blurring artifacts due to off-resonance reconstruction. All $^{13}$C images were acquired at 8 $\times$ 8 $\times$ 20mm and reconstructed at 4 $\times$ 4 $\times$ 20mm. The mean value of the phantom area for each image is normalized by the value of the lactate image at zero frequency offset and plotted in graph (b). Simulation results from Figure 2 are also displayed here for comparison.}
\label{figS:phantom_offres}

\item{Simulation of bSSFP SNR as a function of T2, normalized to SNR is 1 when T2 is 1s. Simulation parameters include T1 = 30s, TR 15.3ms, \# of time points 30, \# of RF pulses 64.}
\label{figS:snr_vs_t2}

\item{Dynamic images and ROI signal curves of a rat kidney slice of the experiments described in Figure 4. Each image is displayed to its own maximum signal to visualize metabolites at all time points. All ROI signals were divided by corresponding noise signals and then divided by the highest value of the pyruvate dynamic curve.}
\label{figS:rat_dynamic}

\item{Dynamic images and ROI signal curves of a TRAMP mouse tumor slice of the experiments described in Figure 5. Each image is displayed to its own maximum signal to visualize metabolites at all time points. All ROI signals were divided by corresponding noise signals and then divided by the highest value of the pyruvate dynamic curve. }
\label{figS:tramp_dynamic}

\item{Dynamic images and ROI signal curves of a human renal tumor slice of the experiments described in Figure 6. Each image is displayed to its own maximum signal to visualize metabolites at all time points. All ROI signals were divided by corresponding noise signals and then divided by the highest value of the pyruvate dynamic curve. }
\label{figS:human_dynamic}

\item{$^{13}$C images of Supporting Figure S2 were displayed without applying demodulation, meaning reconstruction frequency was the same as excitation frequency. This is how reconstruction was performed in all in vivo experiments - reconstruction was always performed at the lactate frequency. This figure validates the combined effects of excitation profiles and blurring artifacts caused by spiral readouts. }
\label{figS:phantom_offres_lacfreq}

\end{enumerate}

\clearpage


\end{document}


\newcommand{\comment}[1]{\todo[inline]{#1}}
\subsection*{Supporting Information}
\renewcommand{\figurename}{Supporting Information Figure}
\renewcommand{\thefigure}{S\arabic{figure}}

\newcommand\rightnote{\normalmarginpar\marginnote}
\newcommand\leftnote{\reversemarginpar\marginnote}
\newcommand\leftnotepar{\reversemarginpar\marginpar}
\newcommand\rightnotepar{\normalmarginpar\marginpar}
\newcommand\partextwidth{\parbox{1\textwidth}}
\newcommand\yellowbox{\colorbox{yellow}}
\newcommand\redtext{\color{red}}

\renewcommand\hl{\textnormal}
\renewcommand\yellowbox{\textnormal}
\renewcommand\partextwidth{\textnormal}
\renewcommand\leftnote{\comment}
\renewcommand\rightnote{\comment}
\renewcommand\rightnotepar{\comment}
\renewcommand\leftnotepar{\comment}
\newpage
\begin{figure} 
\centering  
\includegraphics[width=1\textwidth]{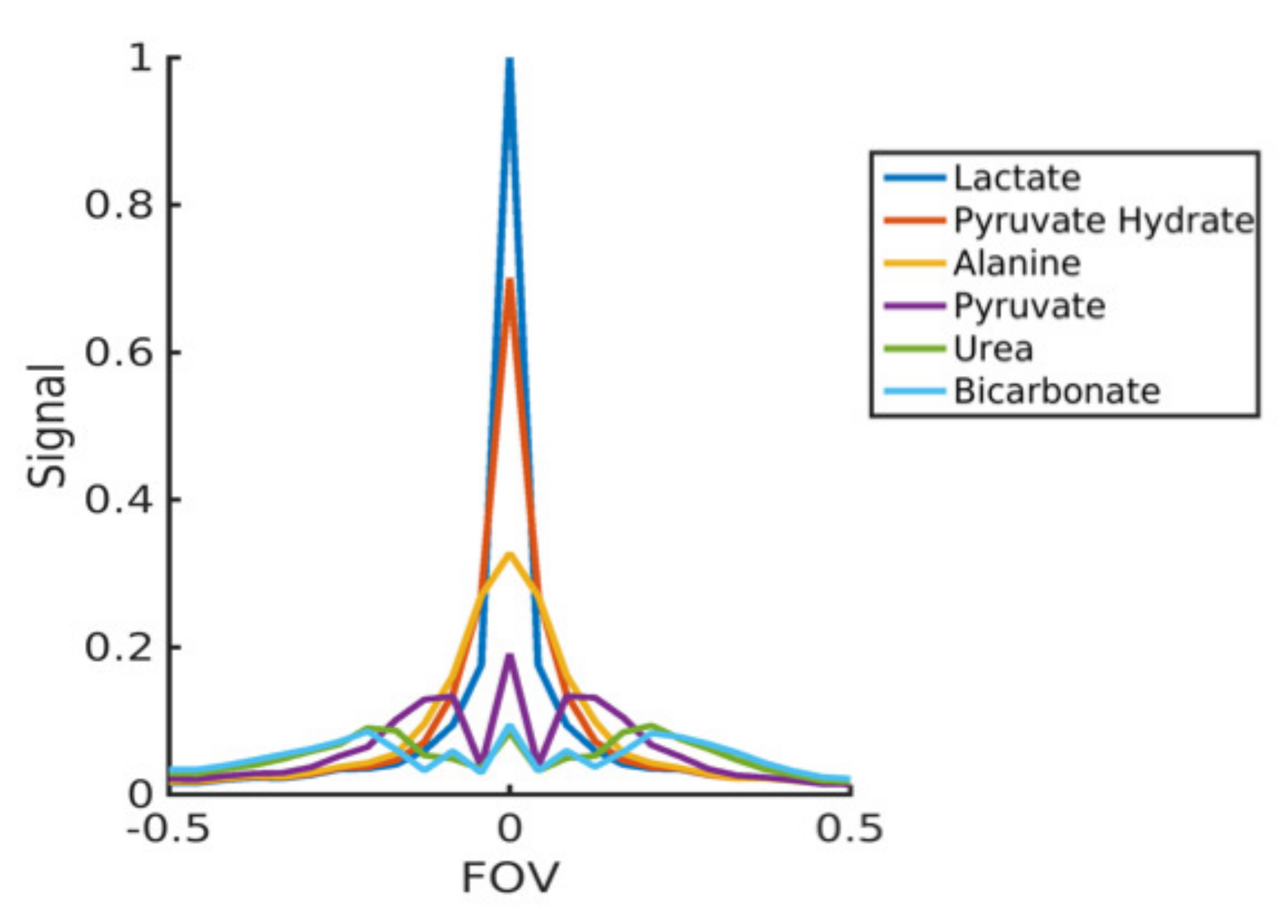}
 \caption
    {
Simulations of off-resonance PSF of the interleaved spiral readouts (4 interleaves, 3.8ms for each interleaf) used in this study. Frequencies of metabolites are Lactate = 0Hz, Pyruvate Hydrate = -128Hz, Alanine = -210Hz, Pyruvate = -395Hz, Urea = -635Hz, Bicarbonate = -717Hz.  
    }
  \label{suppl:sim_offres}
\end{figure}

\begin{figure} 
\centering  
\includegraphics[width=1\textwidth]{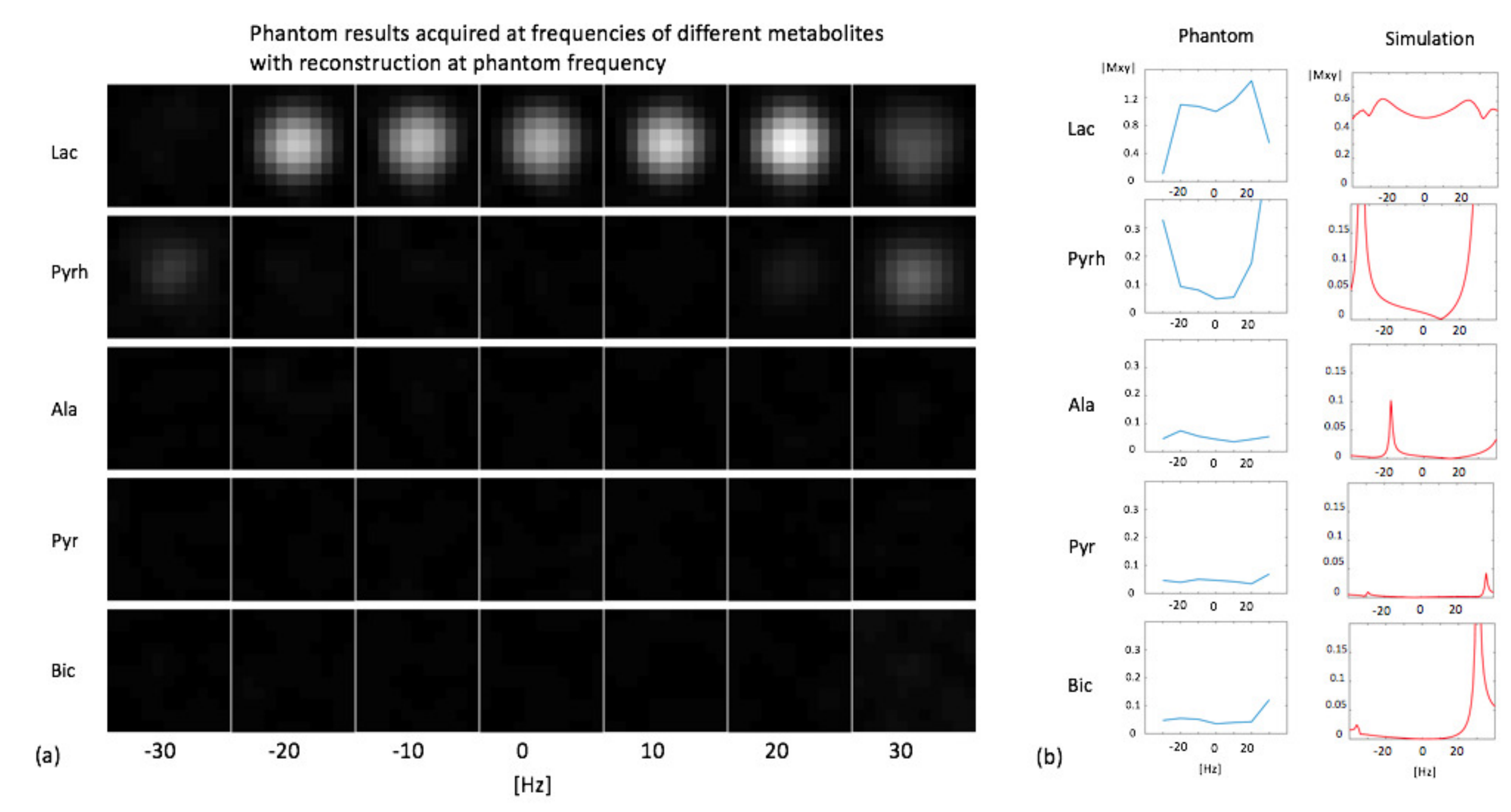}
 \caption
    {
Validation of the MS-3DSSFP sequence on a $^{13}$C-enriched sodium bicarbonate phantom (T1 $\sim$= 26s, T2 $\sim$= 1.5s). $^{13}$C images were acquired with a center frequency offset, both at excitation and at acquisition, by 0Hz, 128Hz, 210Hz, 395Hz, and 717 Hz relative to the phantom frequency, to mimic the images of lactate ("Lac"), pyruvate hydrate ("Pyrh"), alanine ("Ala"), pyruvate ("Pyr") and bicarbonate ("Bic"), respectively. At the frequency of each metabolite, $^{13}$C images were also acquired with small frequency offsets from -30 to 30 Hz with a step of 10Hz. Acquired data were always demodulated to the phantom frequency so that reconstructed images wouldn't be affected by blurring artifacts due to off-resonance reconstruction. All $^{13}$C images were acquired at 8 $\times$ 8 $\times$ 20mm and reconstructed at 4 $\times$ 4 $\times$ 20mm. The mean value of the phantom area for each image is normalized by the value of the lactate image at zero frequency offset and plotted in graph (b). Simulation results from Figure 2 are also displayed here for comparison.
    }
  \label{suppl:phantom_offres}
\end{figure}

\begin{figure} 
\centering  
\includegraphics[width=1\textwidth]{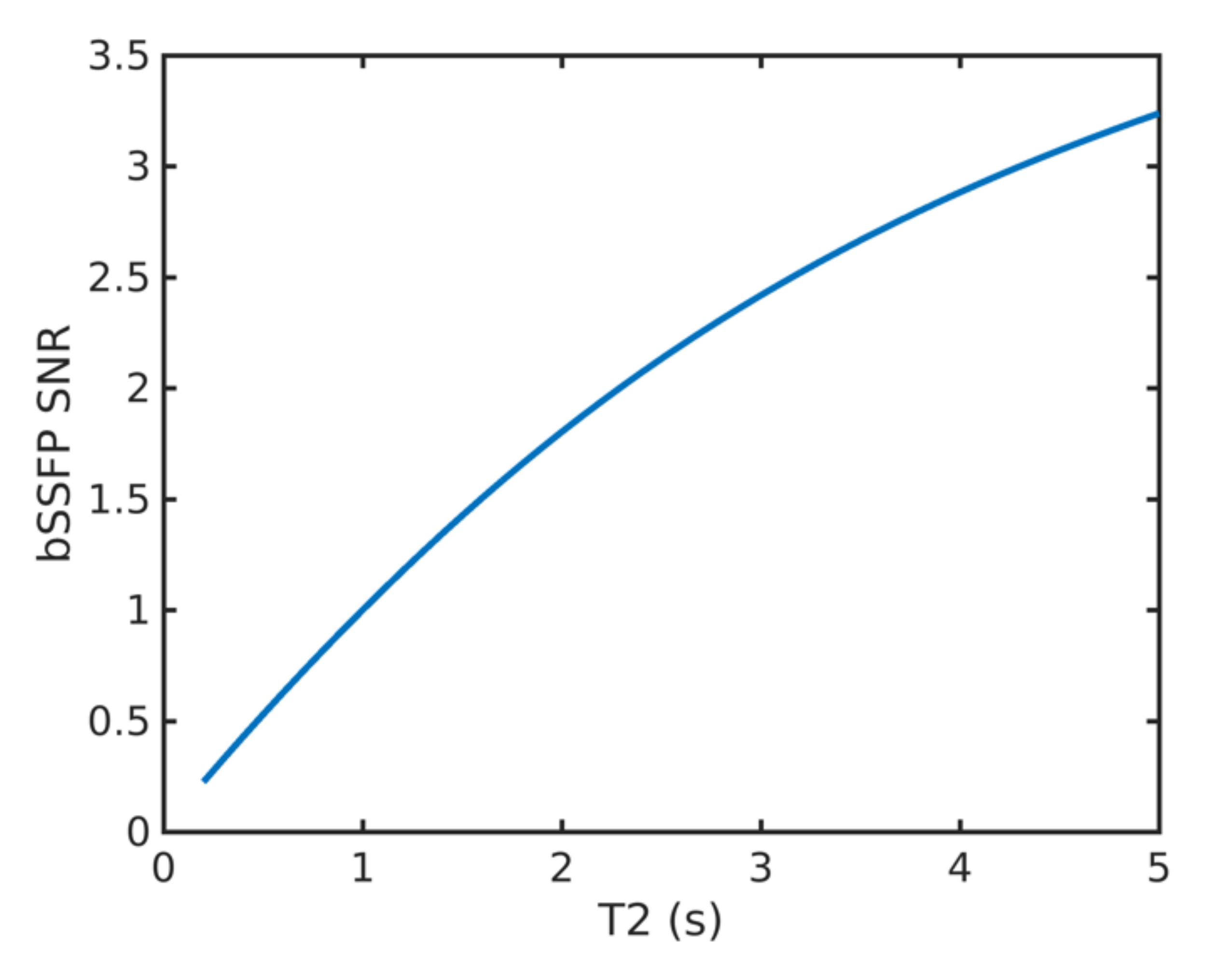}
 \caption
    {
Simulation of bSSFP SNR as a function of T2, normalized to SNR is 1 when T2 is 1s. Simulation parameters include T1 = 30s, TR 15.3ms, \# of time points 30, \# of RF pulses 64.
    }
  \label{suppl:snr_vs_t2}
\end{figure}

\begin{figure} 
\centering  
  \includegraphics[width=1\textwidth]{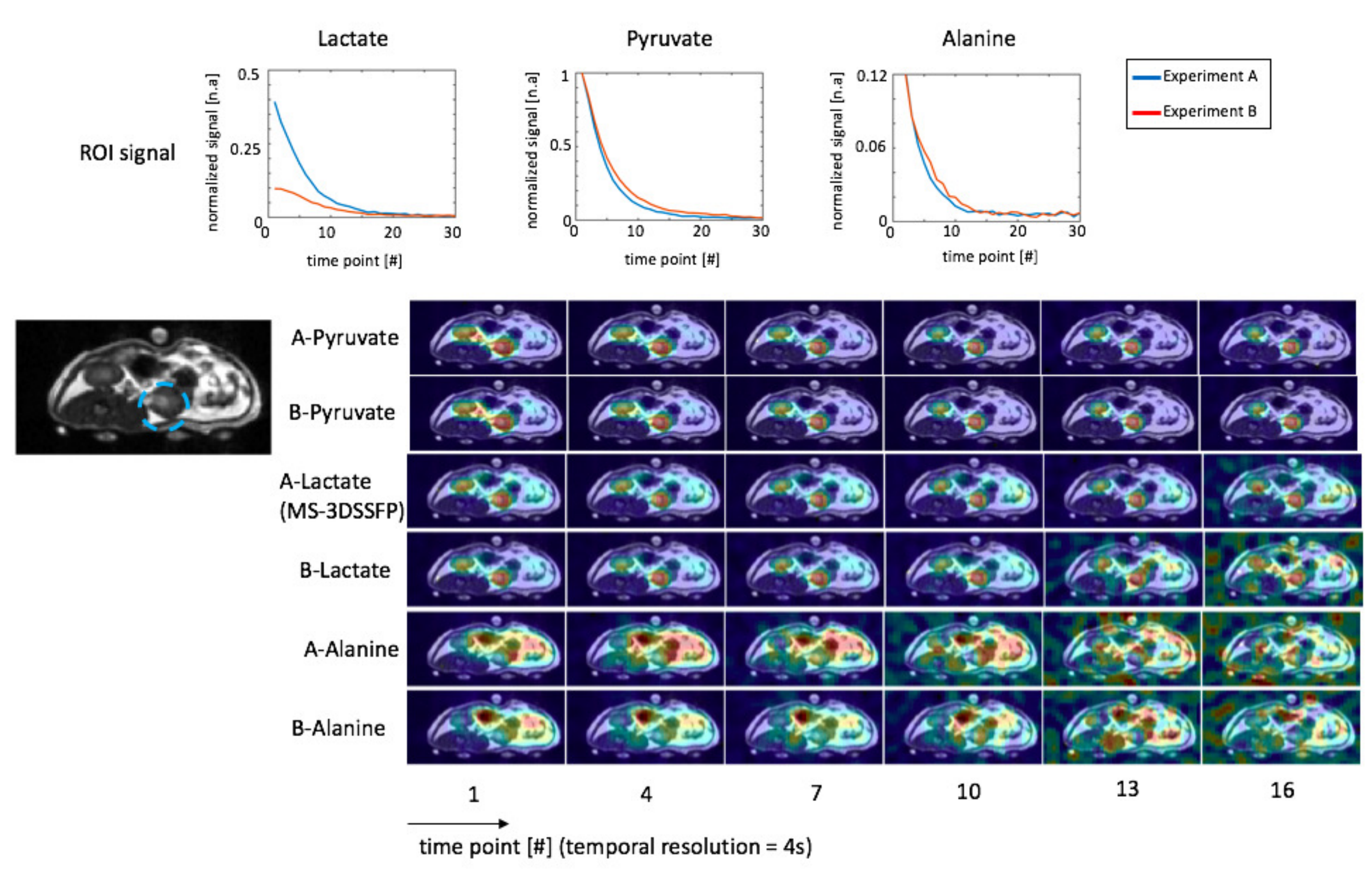}
    \caption
    {
Dynamic images and ROI signal curves of a rat kidney slice of the experiments described in Figure 4. Each image is displayed to its own maximum signal to visualize metabolites at all time points. All ROI signals were divided by corresponding noise signals and then divided by the highest value of the pyruvate dynamic curve.
    }
  \label{suppl:rat_dynamic}    
\end{figure}

\begin{figure} 
\centering  
  \includegraphics[width=1\textwidth]{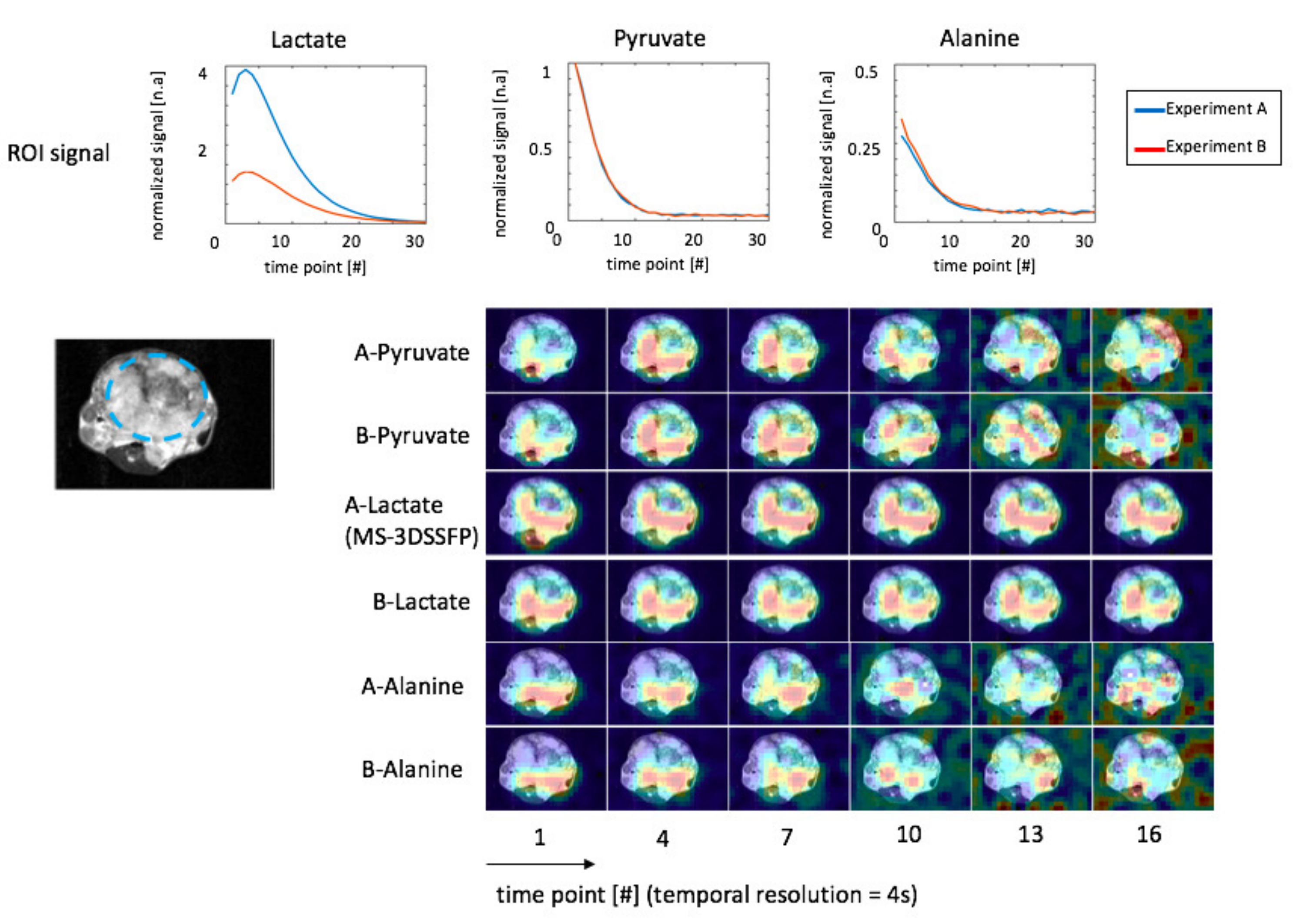}
    \caption
    {
Dynamic images and ROI signal curves of a TRAMP mouse tumor slice of the experiments described in Figure 5. Each image is displayed to its own maximum signal to visualize metabolites at all time points. All ROI signals were divided by corresponding noise signals and then divided by the highest value of the pyruvate dynamic curve.    }
  \label{suppl:tramp_dynamic}    
\end{figure}

\begin{figure} 
\centering  
  \includegraphics[width=1\textwidth]{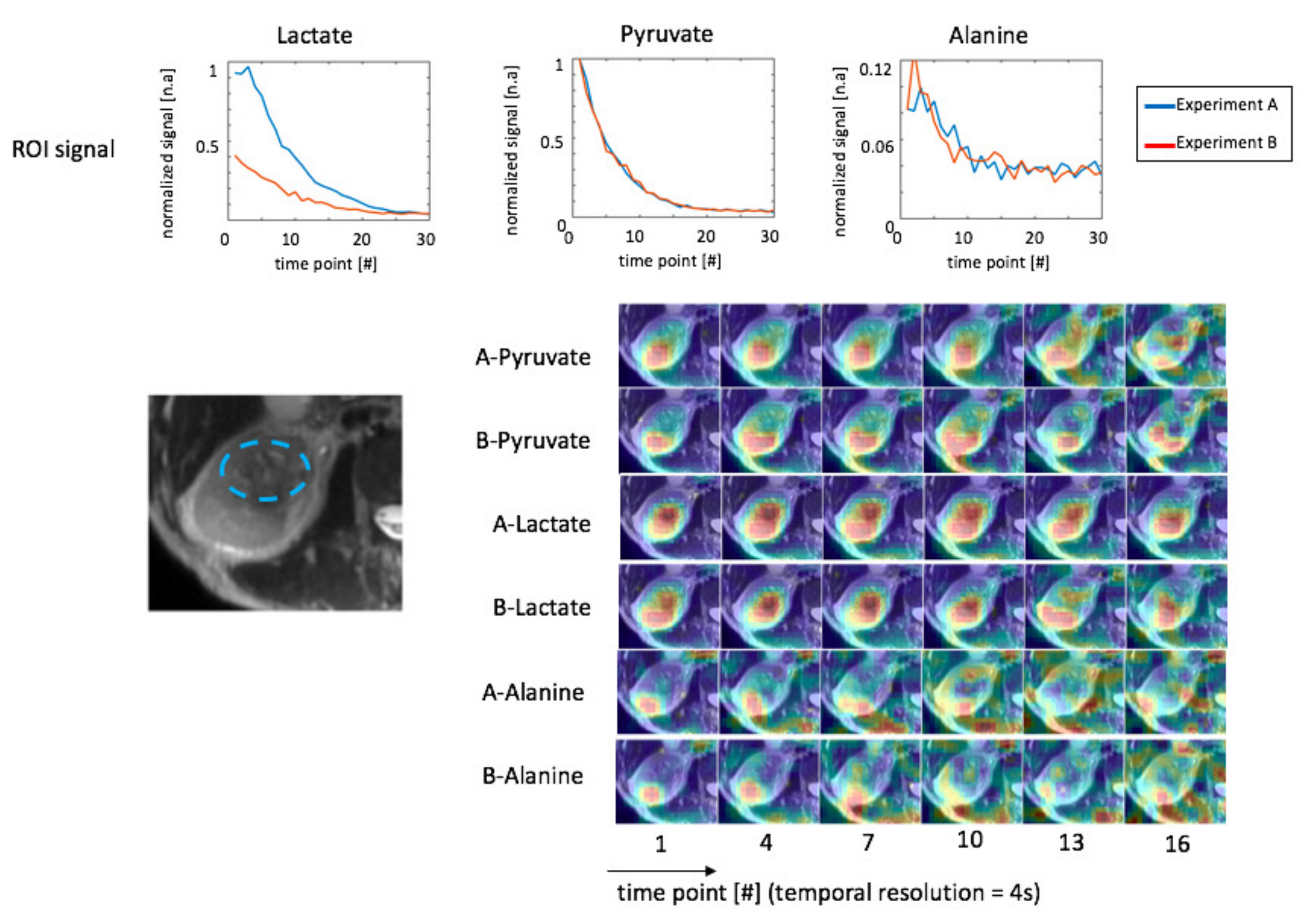}
    \caption
    {
Dynamic images and ROI signal curves of a human renal tumor slice of the experiments described in Figure 6. Each image is displayed to its own maximum signal to visualize metabolites at all time points. All ROI signals were divided by corresponding noise signals and then divided by the highest value of the pyruvate dynamic curve.   }
  \label{suppl:human_dynamic}    
\end{figure}

\begin{figure} 
\centering  
\includegraphics[width=1\textwidth]{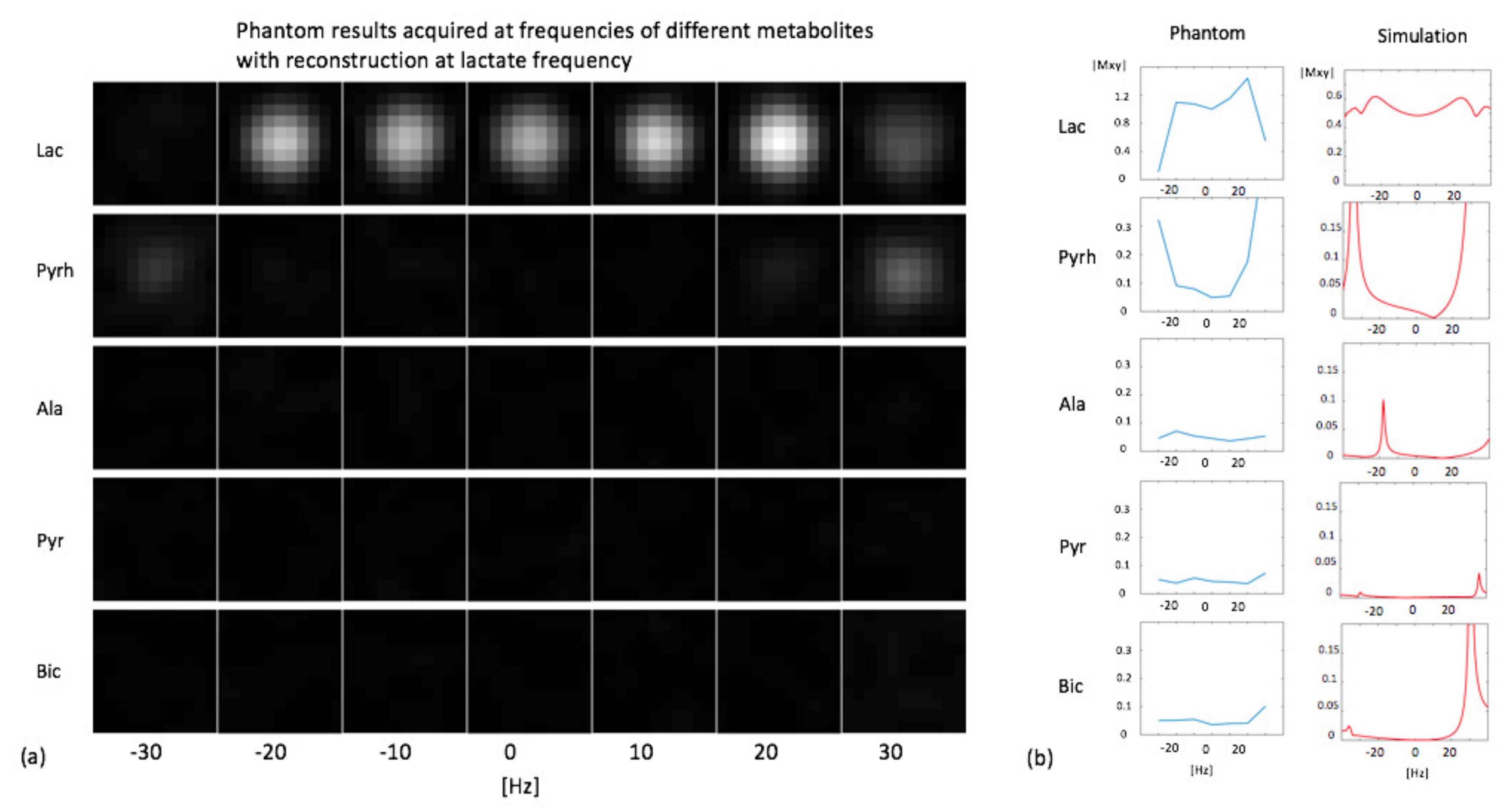}
 \caption
    {
$^{13}$C images of Supporting Figure S2 were displayed without applying demodulation, meaning reconstruction frequency was the same as excitation frequency. This is how reconstruction was performed in all in vivo experiments - reconstruction was always performed at the lactate frequency. This figure validates the combined effects of excitation profiles and blurring artifacts caused by spiral readouts. 
    }
  \label{suppl:phantom_offres_lacfreq}
\end{figure}